\documentclass[a4paper,11pt]{article}
\pdfoutput=1 

\usepackage{jheppub} 

\usepackage[T1]{fontenc} 

\usepackage[utf8]{inputenc}
\usepackage{graphicx}
\usepackage{dcolumn}
\usepackage{bm}

\def\be{\begin{equation}}
\def\ee{\end{equation}}
\def\bea{\begin{eqnarray}}
\def\eea{\end{eqnarray}}
\def\gsim{\ \rlap{\raise 2pt\hbox{$&gt;$}}{\lower 2pt \hbox{$\sim$}}\ }
\def\lsim{\ \rlap{\raise 2pt\hbox{$&lt;$}}{\lower 2pt \hbox{$\sim$}}\ }
\def\dslash{\kern-4pt \not{\hbox{\kern-2pt $\partial$}}}
\def\pslash{\not{\hbox{\kern-2pt p}}}

\def\l{{\rm L}}

\def\bsmumu{b \to s \mu^+ \mu^-}

\def\bsll{b \to s \ell^+ \ell^-}

\def\bra#1{\left\langle #1\right|}
\def\ket#1{\left| #1\right\rangle}

\def \SM{{\rm SM}}

\def \expt{{\rm expt}}

\def \be{\beta}

\def\beq{\begin{equation}}
\def\eeq{\end{equation}}
\def\bea{\begin{eqnarray}}
\def\eea{\end{eqnarray}}
\def\ber{\begin{eqnarray*}}
\def\eer{\end{eqnarray*}}
\def\bwt{\begin{widetext}}
\def\ewt{\end{widetext}}

\def\roughly#1{\mathrel{\raise.3ex\hbox
{$#1$\kern-.75em\lower1ex\hbox{$\sim$}}}}
\def\lsim{\roughly&lt;}
\def\gsim{\roughly&gt;}

\def\order{\lower 1.8ex \hbox{\LARGE\~{}}}

\def\RK{R_K}

\def\RKstar{R_{K^*}}

\usepackage{bm}

\newcommand{\bctaunutau}{b \to c \tau^- {\bar\nu}_\tau}

\def\bra#1{\left\langle #1\right|}
\def\ket#1{\left| #1\right\rangle}

\def \({\left(}
\def \){\right)}
\def \[{\left[}
\def \]{\right]}
\def \l|{\left|}
\def \r|{\right|}

\def \be{\beta}

\def \SM{{\rm SM}}

\def \expt{{\rm expt}}

\def\Z{Z^{\prime}}

\def\bsmumu{ b \to  s \mu^+ \mu^-}

\def\RK{R_K}

\def\RKstar{R_{K^*}}

\def\gsim{\ \rlap{\raise 2pt\hbox{$>$}}{\lower 2pt \hbox{$\sim$}}\ }
\def\lsim{\ \rlap{\raise 2pt\hbox{$<$}}{\lower 2pt \hbox{$\sim$}}\ }
\begin{document}

\preprint{MI-TH-1887}

\title{Neutrino scattering and $B$ anomalies from hidden sector portals }

\author[a, b]{Alakabha Datta,}
\author[c]{Bhaskar~Dutta,}
\author[c]{Shu Liao,}
\author[d]{Danny Marfatia}
\author[c]{and Louis~E.~Strigari}

\affiliation[a]{Department of Physics and Astronomy, 108 Lewis Hall, University of Mississippi, Oxford, MS 38677-1848, USA}
\affiliation[b]{Department of Physics and Astronomy, University of California, 4129 Frederick Reines Hall, Irvine, CA 92697-4575, USA }
\affiliation[c]{Mitchell Institute for Fundamental Physics and Astronomy,
   Department of Physics and Astronomy, Texas A\&M University, 4242 TAMU, College Station, TX 77845, USA}
\affiliation[d]{Department of Physics and Astronomy, University of Hawaii-Manoa, 2505 Correa Road, Honolulu, HI 96822, USA}



\abstract{We examine current constraints on and the future sensitivity to the strength of couplings between quarks and neutrinos in the presence of a form factor generated from loop effects of hidden sector particles that interact with quarks  via new interactions. We consider models associated with either vector or scalar interactions of quarks and leptons generated by hidden sector dynamics. We study constraints on these models using data from coherent elastic neutrino-nucleus scattering and solar neutrino experiments and demonstrate how these new interactions may be discovered by utilizing the recoil spectra. We show that our framework can be naturally extended  to explain the lepton universality violating neutral current $B$ decay anomalies, and that in a model framework the constraints from neutrino scattering can have implications for these anomalies.}

\maketitle


\section{Introduction}
In spite of the success of the standard model (SM) in describing the particle interactions observed in nature, neutrino interactions with matter
are not thoroughly understood.
Many experiments are now making precise measurements of neutrino-nucleus scattering cross sections and neutrino-electron elastic scattering cross sections. The measurements are now precise enough that they are able to probe beyond the SM physics. Recently, the COHERENT experiment at the Spallation Neutron Source (SNS) has measured coherent elastic neutrino-nucleus scattering (CE$\nu$NS) for the first time, finding that the cross section for scattering on CsI is consistent with the SM at approximately $1\sigma$~\citep{Akimov:2017ade}. In addition, measurements by Borexino of the solar neutrino flux, in particular the $^7$Be component, now provide the best measurement of the neutrino-electron elastic scattering cross section at electron recoil energies $\lesssim$ MeV~\citep{Agostini:2017ixy}. 

Because of this plethora of current and future experimental data, it is imperative to consider new theoretical ideas for neutrino interactions in these low energy experiments. Consider axial-vector interactions between quarks and neutrinos mediated by a new $Z^{\prime}$ boson. We write the interaction of the quarks with the $Z^{\prime}$ as,
\beq
{\cal{ L}}_{ q' q'} = \, \bar{q'} \hat{\gamma^{\mu}}\left[  P_LF_L(q^2) +  P_R F_R(q^2) \right]  q' \, Z^\prime_{\mu} ~,
\label{qqZ}
\eeq
where  $ \hat{\gamma^{\mu}}=  \left[ \gamma_\mu - \frac{\gamma\cdot q\, q^{\mu}}{q^2} \right]$ and $q'$ are SM quark fields. The interaction of the $Z^{\prime}$ with the leptons has an analogous expression.
In the interaction above, the contribution from the $q^{\mu}$ part may be suppressed by small masses or vanish from current conservation. 
Form factors proportional to  $ \sigma^{\mu \nu} q_\nu$ are possible, but  will be suppressed by some hadronic scale; we do not investigate these in this paper.

 The form factor $F(q^2)$ can be unity when $q'$ couples directly to $Z'$. However, in many models (especially the models with low scale hidden sectors), $q'$ may  couple to  $Z'$ via a loop containing DM/hidden sector particles. In such a scenario we expect $F(q^2)\sim q^2/\Lambda^2$ where $\Lambda$ is the scale in the DM sector associated with the  mass of the mediator particle that generates the quark-DM interactions, $\bar{q}q\bar{\chi}\chi$. As long as $\Lambda$ is greater than $q_{\rm max}$ for these scattering experiments, $F(q^2)\sim q^2/\Lambda^2$ appears in the scattering amplitude. In this paper, we investigate these new form factors at CE$\nu$NS  (COHERENT and reactor based) and Borexino experiments for vector and scalar mediators.  We extend our framework to study the neutral current $B$ decay anomalies in the $R_K$ and $R_{K^*}$ measurements. We show how in a model framework measurements from
 neutrino scattering may have implications for the $B$ anomalies.

The paper is organized as follows. In Section II, we discuss models for form factors and their dependence on the new physics scale. In Sections~III
and IV  we discuss the  effects of these new form factors at CE$\nu$NS and Borexino experiments, respectively.  In Section V we discuss the $B$ anomalies and the implications of neutrino scattering experiments on their explanations. We conclude in Section~VI.



\section{Form factors}
\subsection{$\Z$ Model}

\begin{figure}[t]
	\includegraphics[width=7.5cm]{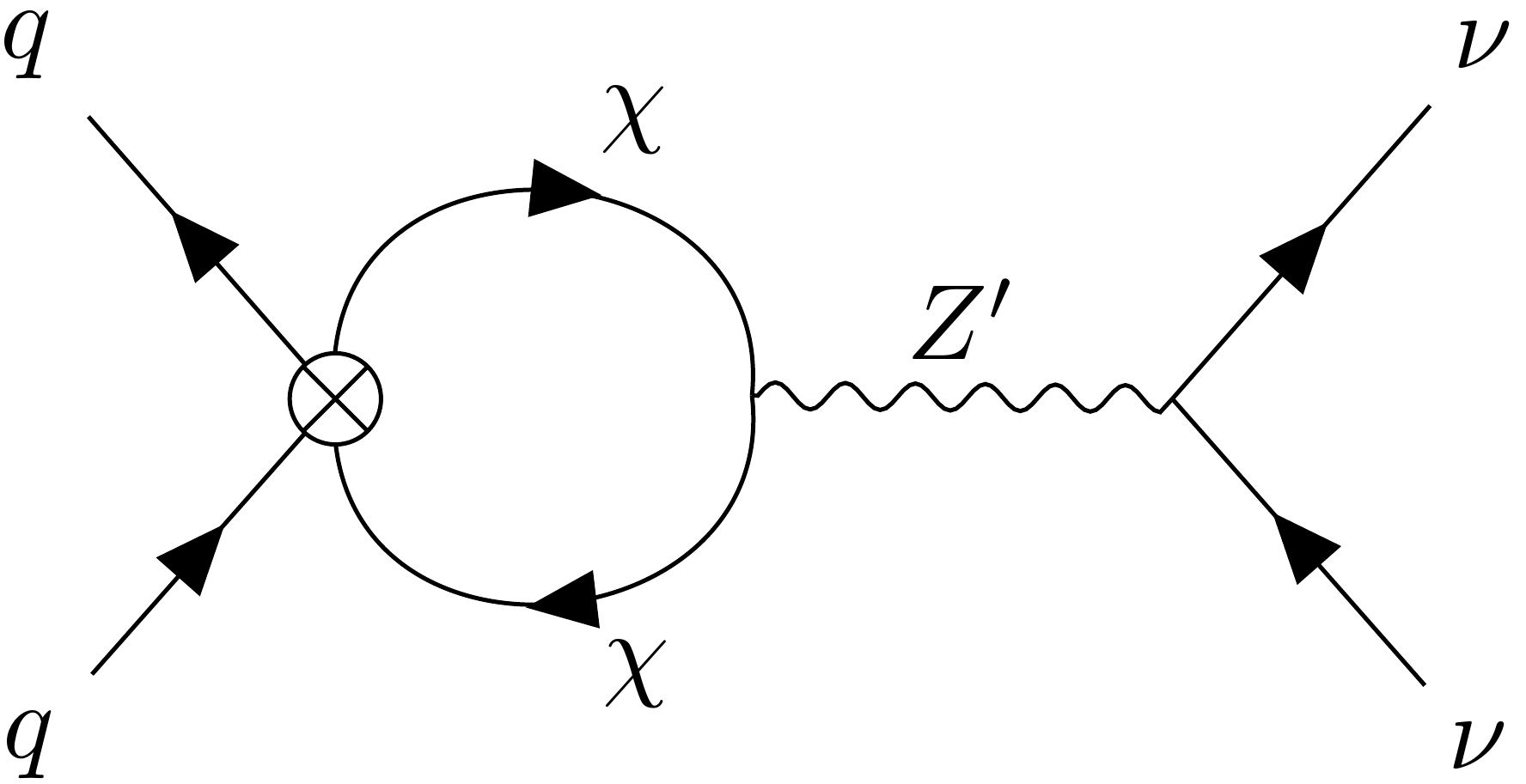}\ \ \ \ \includegraphics[width=7.5cm]{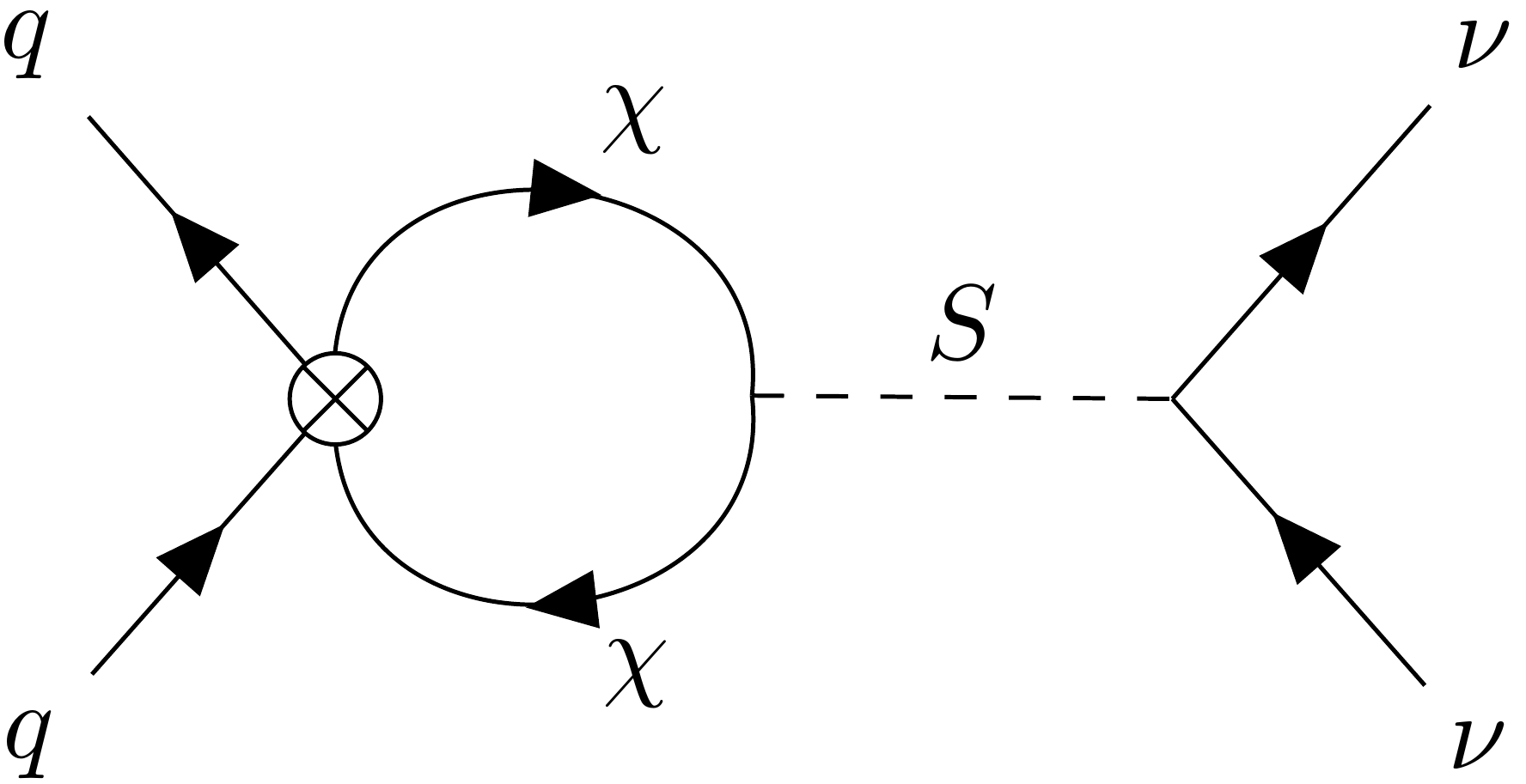}
	\caption{Feynman diagram for $Z^\prime$ model (left) and $S$ model (right).\label{fig:feynman}}
\end{figure}

Consider the following Lagrangian at
  low energy~\cite{SZMEV}:
  \bea
\cal{L} & =&\frac{g}{\Lambda^2} \bar{q'} \gamma^{\mu} P_{L,R} q' \bar{\chi}\gamma_\mu ( 1 \pm \gamma_5) \chi +i  \bar{\chi} \gamma^\nu \left[ \partial_\nu -i g_{\chi}    Z^{\prime \nu}\right]\chi - m_{\chi} \bar{\chi}{\chi} + \frac{1}{2}m_{Z^{\prime}}^2 Z_{\mu}^{\prime}Z^{ \prime \mu} \nonumber\\
&=& H_{eff} + J_{\mu, \chi} Z^{\prime, \mu} +i  \bar{\chi} \gamma^\nu  \partial_\nu \chi - m_{\chi} \bar{\chi}{\chi} + \frac{1}{2}m_{Z^{\prime}}^2 Z_{\mu}^{\prime}Z^{ \prime \mu}\,,
\label{lagZprime}
\eea
where $\chi$ is a hidden sector fermion field with mass $m_{\chi}$. The first term in the Lagrangian represents an effective coupling between the $q$ and $\chi$ fields that 
might arise through the exchange of a heavy mediator of mass $ \sim \Lambda$, with $\Lambda \gg E$, where $E$ is the energy scale of the process (see for example~\cite{Goodman:2010ku,Bai:2010hh,Fan:2010gt}). The hidden sector fields $\chi$ couple directly to $\Z$ through the vector portal and so
in our framework there are two mediators; see e.g.,~\cite{Elor:2018xku}. We further assume that the neutrinos are charged under the $\Z$ $U(1)$ and so there is a direct coupling of the neutrinos to $\Z$. Although there is no direct coupling between the  quarks and $\Z$ field, 
 a $\chi$-loop-induced $ \bar{q} q \Z$ effective vertex (as in Eq.~\ref{qqZ})  will be generated with a $q^2$ dependent coupling, which can be represented by a higher dimensional operator,
 \beq
 {\cal{L}}_{HD} = \frac{g_{L,R }}{\Lambda^2}  \bar{q'} \gamma^{\mu} P_{L,R} q' \partial^\nu Z^{\prime}_{\mu \nu}\,,
 \label{hd}
 \eeq
 where $Z^{\prime}_{\mu \nu} $ is the $\Z$ field tensor; see Fig. \ref{fig:feynman}.
This higher dimensional operator may be considered to be the bare term of the Lagrangian.

 The form factors in Eq.~(\ref{qqZ}) are then given by
\bea
\bar{q'} \hat{\gamma^{\mu}}\left[  P_LF_L(q^2) +  P_R F_R(q^2) \right]  q'      
&= & \frac{1}{\Lambda^2} \bra{q'} \int d^4x e^{iq.x}T \left[ J_{\mu, \chi}(x) H_{eff}(0) \right]\ket{q'}. \
\label{FFdef}
\eea
The most general structure for the form factors  from the current conservation of $J_{\mu,\chi}$ is
\bea
F_L(q^2) & = & \frac{q^2}{\Lambda^2} A_L(q^2)\,,
 \nonumber\\
F_R(q^2) & = & \frac{q^2}{\Lambda^2}  A_R(q^2)\,.
\label{FFgeneral}
\eea
Since the matrix element in Eq.~(\ref{FFdef}) is finite as $q^2 \to 0$, $A_{L,R}(q^2)$ is finite as $q^2 \to 0$.
Now from Eqs.~(\ref{hd}) and ~(\ref{FFdef}), we can estimate the form factor in the 1-loop approximation~\cite{Datta:2013kja}.
We get
\bea
F_{L,R}(q^2) & = &\frac{ q^2}{\Lambda^2} \left[ g_{L,R} +gg_\chi\frac{1}{ (2 \pi)^4}  \int_0^1 dx \  8x(1-x) \int d^4 l \frac{ 1}{( l^2- \Delta + i \epsilon)^2}\right], \
\label{loop0}
\eea
where $ \Delta= m_{\chi}^2- q^2 x(1-x) $.  Introducing a cut-off $\Lambda_c$ to regulate the divergent integral we can write
\bea
F_{L,R}(q^2)  & = & \frac{q^2}{\Lambda^2}\left[ g_{L,R} + \frac{g g_{\chi} }{ 2 \pi^2}  \int_0^1 dx \  x(1-x)\ln\frac{ x\Lambda_c^2}{\Delta}\right], \nonumber\\
&= &   \frac{q^2}{\Lambda^2} g_{L,R}(q^2) .\
\label{loop_regulate}
\eea
where $g$ and $g_{\chi}$ are the bare coupling constants. We rewrite $g_{L,R}(q^2) $ as
\bea
g_{L,R}(q^2) & = & g_{L,R} ( q^2_{max}) +  \frac{g g_{\chi} }{ 2 \pi^2}  \int_0^1 dx \  x(1-x)\ln\frac{ \Delta_{max}}{\Delta}, \
\label{gLR}
\eea
where  $ \Delta_{max}= m_{\chi}^2- q^2_{max} x(1-x) $ and $q^2_{max}$ is the maximum momentum transfer squared.

Note that this  is a rough estimate as  we have calculated only the leading term in Eq.~(\ref{FFdef}). However, the general structure of Eq.~(\ref{FFgeneral}) still holds as it follows from vector current conservation.
 As a rough estimate, assuming all terms in Eq.~(\ref{gLR}) to be of the same size, we can write,
 \bea
g_{L,R}(q^2) & \sim &   \frac{g g_{\chi} }{ 2 \pi^2}  \int_0^1 dx \  x(1-x)\ln\frac{ \Delta_{max}}{\Delta}. \
\label{gLRestimate}
\eea 
 
For our purposes, the most important part of Eq.~(\ref{loop0}) is the $q^2/\Lambda^2$ factor which is not present if either the tree level (first term)  or the loop contribution (second term) is absent; the latter is Eq.~(\ref{gLRestimate}). We show both tree and loop contributions since each term may not be sizable in a given model.  The relative strength between  the tree and the loop contributions does not affect our analysis.

 Although the Lagrangian in Eq.~(\ref{lagZprime}) contains both $g_L$ and $g_R$ terms, only $g_L + g_R \equiv g_v$ contributes to $\nu$-nucleus coherent scattering; $g_L - g_R \equiv g_a$ does not impact the scattering process. This is because the vector charge of the nucleus is proportional to the number of nucleons, $Q_v = Zg^p_v + Ng^n_v$ while the axial vector couplings are proportional to the spin, $Q_a = g^p_a\left<S_p\right> + g^n_a\left<S_n\right>$ which is much smaller than $Q_v$. 
  
\subsection{$S$ Model}

As for the $\Z$ case the  form factors for a scalar mediator $S$ can be written as 
\bea
\bar{q'} \left[ P_LS_L(q^2) +  P_R S_R(q^2) \right]  q'      & = &  
=\frac{1}{\Lambda^2} \bra{q'} \int d^4x e^{iq.x}T \left[ J_{ \chi}(x) H_{eff}(0) \right]\ket{q'}, \
\label{FFdefS}
\eea
where
\bea
H_{eff} & = & \frac{g}{\Lambda^2} \bar{q}  P_{L,R} q \bar{\chi}\ ( 1 \pm \gamma_5) \chi\,, \nonumber\\
J_{\chi} & = &  \bar{\chi} \left [ g_{\chi} P_L + g_{\chi}^\prime P_R \right] \chi\,.
\eea
In this case we cannot use current conservation. By calculating the leading loop contribution we find,
\bea
S_{L,R}(q^2) & = & gg_\chi\frac{1}{ (2 \pi)^4}  \int_0^1 dx \  12 x(1-x)  \frac{  \Delta}{\Lambda^2}   \int d^4 l \frac{ 1}{( l^2- \Delta + i \epsilon)^2}\,.
\label{loop}
\eea
Again introducing a cut-off $\Lambda_c$ we can write,
\bea
S_{L,R}(q^2)  & \sim & \frac{ 3g g_\chi}{ 4 \pi^2}  \int_0^1 dx \  x(1-x)\frac{ \left[ m_\chi^2 - q^2 x(1-x) \right]}{\Lambda^2}\ln\frac{ x\Lambda_c^2}{\Delta}\,.
\label{loopS}
\eea
For $m_{\chi}^2 \ll q^2$, $S_{L,R}(q^2) \sim q^2$.  Unlike the  $\Z$ case the form factor is sensitive to the $\chi$ mass and so the hidden sector dynamics can be probed in low energy scattering.

A scalar coupling has a larger effect on the coherent scattering rate $\sim O\left(E_R\right)$, compared to the rate from a pseudoscalar coupling, $\frac{1}{8\pi p^2}\frac{g_\nu^2 g_q^2}{\left(2m_NE_R+m_s^2\right)^2}E_R^2m_N\sim O\left(E_R^2\right)$. 

The scalar interactions of quarks with dark matter
$ \sim \frac{g}{\Lambda^2}\bar{q}P_{L,R} q \bar{\chi} P_{L,R} \chi$  could arise from a higher dimensional operator
$\frac{g}{\Lambda^3} H \bar{q}P_{L,R} q \bar{\chi} P_{L,R} \chi$. With the assumption of minimal flavor violation (MFV) this interaction has the form
 $ \sim \frac{m_q}{\Lambda^3}  \bar{q}P_{L,R} q \bar{\chi} P_{L,R} \chi$~\cite{Goodman:2010ku}.
 
 It is worth considering how the phenomenology changes if one uses MFV to constrain the scalar interactions. Without MFV
 the matrix element that appears in the amplitude is $ M \sim  \bra{N} \bar{q}q \ket{N}$ while with MFV we need
 $M'  \sim \frac{1}{\Lambda}  \bra{N} m_q\bar{q}q \ket{N}$.  If the heavy $c, b, t$ quarks are involved they can be integrated out and their contributions
 replaced by a gluonic term~\cite{Cerdeno:2016sfi, DelNobile:2013sia}. If contributions arise only from the light quarks then
 ${M' \over M} \sim {m_q \over \Lambda}$, and for $\Lambda \gg m_q$  the matrix element for the MFV case is suppressed. Hence, the bounds on $\Lambda$ from coherent scattering will be weaker.
 
 When we discuss the $B$ anomalies below, we will require that dark matter couple to quarks of all generations or at least to all generations of down quarks. Consequently, the heavy quarks could contribute to coherent scattering via the gluonic terms. However, for the $B$ anomalies we need a flavor changing coupling $g_{bs}$ which involves a mixing angle. In other words, the $B$ anomalies do not completely fix the coupling $g_q$, of the heavy quarks to dark matter. The  heavy quark contribution to the matrix element $M$ 
 goes as $ \sim  g_q {m_N \over m_q}$ while for
$M'$ it goes as $ \sim g_q   {m_N \over \Lambda}$, where $m_N$ is the nucleon mass.

Note that the scalar coupling to the neutrinos  may originate from the  lepton number violating interaction, $\phi \bar{\nu}^c_L\nu_L$, or from the lepton number conserving interaction, $\phi\bar{\nu}_R\nu_L$, where  $\phi$ is a SM scalar singlet. 
If $\phi$ is a pseudoscalar, then the associated  coupling is not constrained by coherent scattering but is constrained by
$\pi^0 \to \bar{\nu}\nu$ \cite{Tanabashi:2018oca} to be smaller than 
 $10^{-5}$.

Finally, depending on whether or not right-handed neutrinos are present, we can write neutrino-quark interaction terms as 
$g^\prime /m^{\prime 2}\bar{\nu}_R\nu_L\bar{q} q$ or  $g^\prime /m^{\prime 2}\bar{\nu}_L^c\nu_L\bar{q} q$. If the second term is present, then a Majorana mass term arises from the quark condensate, i.e., $m_\nu=g^\prime /{(m^\prime/\mathrm{GeV})}^2 \bar{q}q =g^\prime /{(m^\prime/\mathrm{GeV})}^2 10^7$ eV  with $\bar{q} q=8 \pi/\sqrt{3} f_\pi^3$ as calculated in the Nambu-Jona-Lasinio model~\cite{Nambu:1961tp}. Since the heaviest neutrino mass is $\sim 0.1$~eV, we find $g^\prime /{(m^\prime/\mathrm{GeV})}^2 \leq 10^{-8} $. However, $g^\prime$  is constrained by COHERENT to be smaller than $10^{-10} \ (10^{-8})$ for $m^{\prime}=10$~MeV (1~GeV)~\cite{us}.

\section{Scattering cross sections}

For the vector model we can write the effective interaction as
\begin{equation}
	\mathcal{L}_{\mathrm{BSM}} = -2\sqrt{2}G_F\bar{\nu_L}\gamma^\mu\nu_L\bar{f}\gamma_\mu f \frac{g\prime F\left(q^2,\Lambda^2\right)}{q^2+m^{\prime2}}\frac{1}{2\sqrt{2}G_F}\,,
\end{equation}
where $F\left(q^2,\Lambda^2\right)=\frac{q^2}{\Lambda^2}$. 
Since the vector interaction has the same structure as in the SM, its contribution can interfere with the SM contribution.

The neutrino-electron differential cross section can be written as
\begin{equation}
	\frac{d\sigma}{dE_R} = \frac{2}{\pi}G_F^2 m_e \left[\left(\epsilon^L\right)^2+\left(\epsilon^R\right)^2\left(1-\frac{E_R}{E_\nu}\right)^2-\epsilon^L\epsilon^R\frac{m_eE_R}{E_\nu^2}\right]\,,
\end{equation}
where $\epsilon^L=\frac{1}{2}+\sin^2\theta_w+\frac{g\prime F\left(q^2,\Lambda^2\right)}{q^2+m^{\prime2}}\frac{1}{2\sqrt{2}G_F}$ and $\epsilon^R=\sin^2\theta_w+\frac{g\prime F\left(q^2,\Lambda^2\right)}{q^2+m^{\prime2}}\frac{1}{2\sqrt{2}G_F}$ for the electron neutrino, and $\epsilon^L=-\frac{1}{2}+\sin^2\theta_w+\frac{g\prime F\left(q^2,\Lambda^2\right)}{q^2+m^{\prime2}}\frac{1}{2\sqrt{2}G_F}$ and $\epsilon^R=\sin^2\theta_w+\frac{g\prime F\left(q^2,\Lambda^2\right)}{q^2+m^{\prime2}}\frac{1}{2\sqrt{2}G_F}$ for the muon or tau neutrino.

The neutrino-nucleus differential cross section is
\begin{equation}
	\frac{d\sigma}{dE_R} = \frac{2}{\pi}G_F^2 m_N Q_V^2 \left[1-\frac{m_NE_R}{E_\nu^2}+\left(1-\frac{E_R}{E_\nu}\right)^2\right]F_{nucl}\left(q^2\right)\,,
\end{equation}
where $F_{nucl}\left(q^2\right)$ is the nuclear form factor, the ``weak charge" $Q_V$ is given by
\begin{equation}
	Q_V = \frac{1}{2}\left[Z\left(\frac{1}{2}-2\sin^2\theta_w\right)+N\left(-\frac{1}{2}\right)+3\left(Z+N\right)\frac{g\prime F\left(q^2,\Lambda^2\right)}{q^2+m^{\prime2}}\frac{1}{2\sqrt{2}G_F}\right]\,,
\end{equation}
We have assumed the $\Z$ couplings are the same for the up and down quark.

On the other hand, for the scalar model, the effective interaction is
\begin{equation}
	\mathcal{L}_\mathrm{BSM} = \bar{\nu}^c_L\nu_L \bar{f}f \frac{g\prime F\left(q^2,\Lambda^2\right)}{q^2+m^{\prime2}}\,,
\end{equation}
which does not interfere with the SM. We can write its contribution to the differential cross section as
\begin{equation}
	\frac{d\sigma}{dE_R}=\frac{d\sigma}{dE_R}\rvert_{SM} + \frac{1}{4\pi}\left(\frac{g\prime F\left(q^2,\Lambda^2\right)}{q^2+m^{\prime2}}\right)^2 \frac{E_Rm_e^2}{E_\nu^2}\,,
\end{equation}
for electron scattering, and
\begin{equation}
	\frac{d\sigma}{dE_R}=\frac{d\sigma}{dE_R}\rvert_{SM} + \frac{1}{4\pi}\left(\frac{\left(15.1Z+14N\right)g\prime F\left(q^2,\Lambda^2\right)}{q^2+m^{\prime2}}\right)^2 \frac{E_Rm_e^2}{E_\nu^2}\,,
\end{equation}
for nucleus scattering~\cite{Cerdeno:2016sfi}.
Here $Z$ ($N$) corresponds to the number of protons (neutrons).

\section{Experimental bounds} 
This section describes the experiments that we use to bound the aforementioned models. We begin by discussing accelerator and reactor experiments, and then discuss solar neutrino experiments. 

\subsection{Accelerators and reactors} 

To evaluate current and future constraints from accelerator and reactor CE$\nu$NS experiments, we use a $\chi^2$ analysis to calculate bounds on the coupling at the $2\sigma$ confidence level. 
To take into account reactor and accelerator neutrino flux uncertainties, we introduce a nuisance parameter $\alpha$ and an uncertainty on the signal of $\sigma_\alpha$. 
We define
\begin{equation}
  \chi^{2}=\sum_{\mathrm{bins}}\left[\frac{N_{obs}^i-\left(1+\alpha\right)N_{th}^i}{\sigma_{\rm {stat}}^i}\right]^2+\left(\frac{\alpha}{\sigma_{\alpha}}\right)^{2}\,,
\label{chi0}
\end{equation}
where $N_{obs}^i$ ($N_{th}^i$) is the observed (predicted) number of events per bin in a current measurement, $\sigma_\alpha=0.28$ and $\sigma_{\rm {stat}}^i$ is the statistical uncertainty which can be extracted from Ref.~\cite{Akimov:2017ade}. For future measurements with multiple detectors we define (with indices suppressed),
\begin{equation}
  \chi^{2}=\sum_{\mathrm{bins,\,detectors}}\frac{\left(N_{SM}-\left(1+\alpha\right)N_{th}\right)^{2}}{N_{bg}+N_{SM}}+\left(\frac{\alpha}{\sigma_{\alpha}}\right)^{2}\,,
\label{chi}
\end{equation}
where $N_{SM}$ is the expected number of events in the SM for a future measurement and  $N_{bg}$ is the expected number of  background events, which we assume to be known precisely. 
Here we estimate $\sigma_\alpha=0.1$ for future measurement.

The current COHERENT experiment has a threshold $4.25$ keV~\cite{Akimov:2017ade}. For the future projected measurements we assume a threshold of 100~eV for Ge and Si reactor experiments~\cite{aguilar2016connie, li2016neutrino, agnolet2017background}, and 2~keV for NaI and Ar with COHERENT~\cite{Akimov:2018ghi}. For reactor neutrinos we take a background of 1~dru (Ge and Si), and for accelerator neutrino data we take a background of $5\times10^{-3}$~dru (CsI, NaI and Ar)~\cite{Akimov:2017ade}.
Here the unit dru stands for differential rate unit, equal to $\mathrm{event}/\left(\mathrm{keV \cdot kg \cdot day}\right)$.
The COHERENT experiment has an energy dependent efficiency. 
We applied the efficiency function from \cite{Akimov:2017ade} to all the detectors in the COHERENT experiment.
We take the reactor neutrino flux to be that of a 1~MW reactor at $\sim 1$~m from the core (which yields a the total flux of $1.5\times10^{12}~\mathrm{cm^2/s}$), and the antineutrino fission spectrum at various sites from Ref.~\cite{an2017improved}. The accelerator neutrino flux at SNS is $4.29\times10^9~\mathrm{cm^2/s}$~\cite{Akimov:2017ade}.

\begin{figure}[t]
\label{coherent_vector}
  \centering\includegraphics[width=12.5cm]{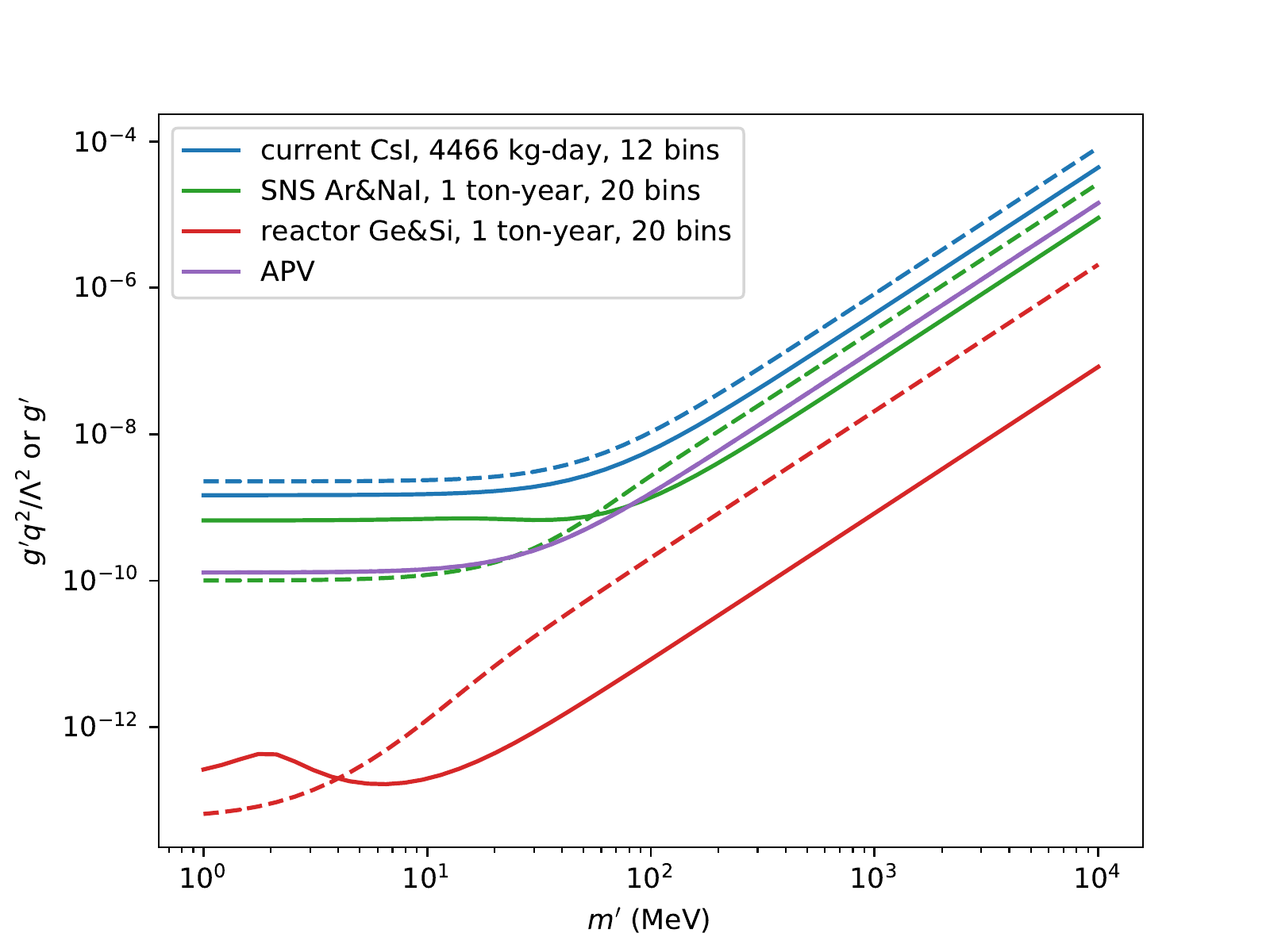}~~\\
   \caption{Current and projected 2$\sigma$ bounds on a vector mediator with $F\left(q^2\right)\sim q^2$ as a function of the mediator mass. Dashed lines show the limits without a form factor. Here $q_0=50$~MeV for COHERENT, and $q_0=30$~MeV for reactor experiments.\label{fig:vector}}
 \end{figure}

\begin{figure}[t]
\label{coherent_scalar}
  \centering\includegraphics[width=12.5cm]{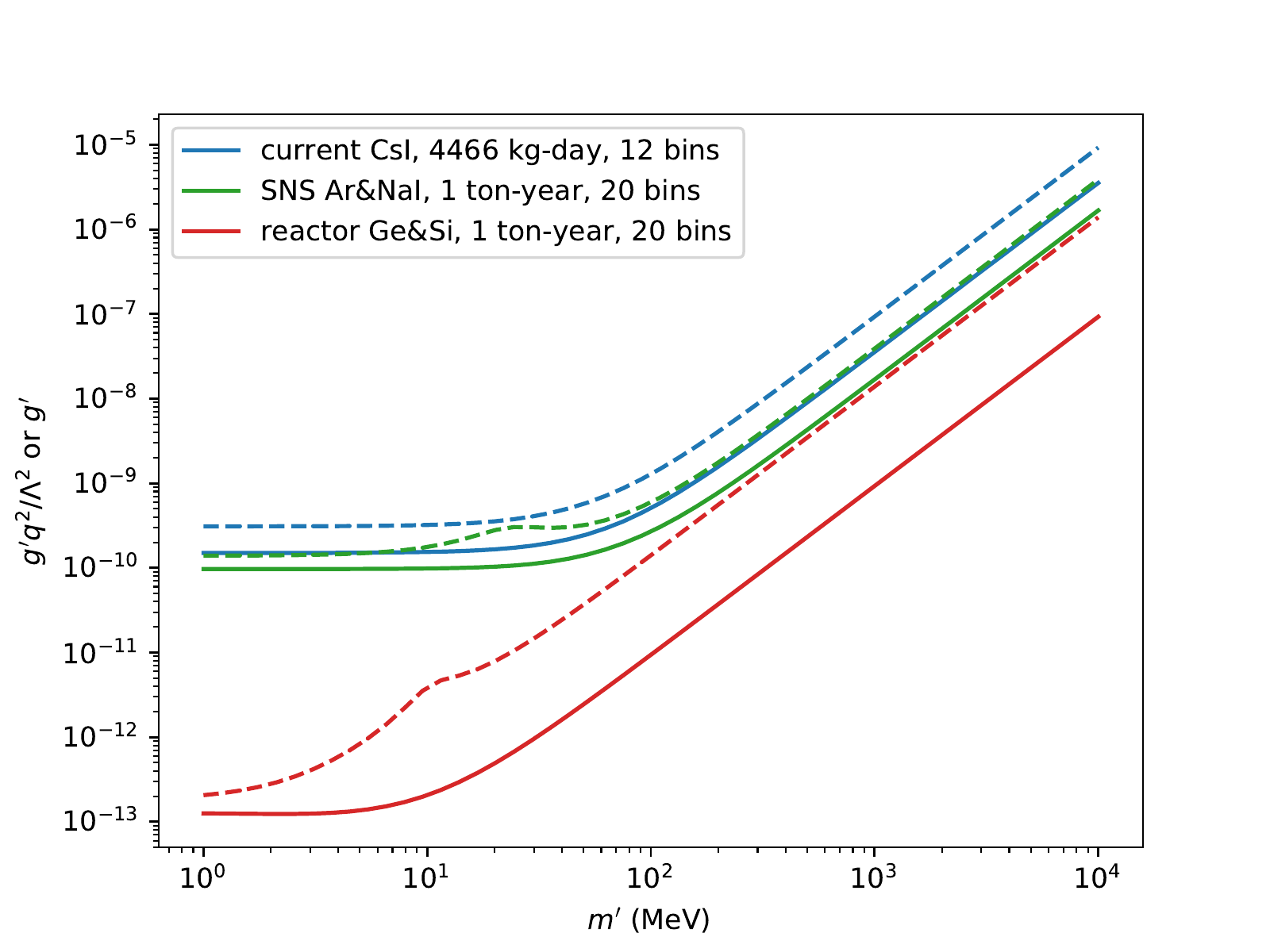}~~\\
   \caption{Current and projected 2$\sigma$ bounds on a scalar mediator with $F\left(q^2\right)\sim q^2$ as a function of the mediator mass. Dashed lines show the limits without a form factor. Here $q_0=50$~MeV for COHERENT, and $q_0=30$~MeV for reactor experiments.\label{fig:scalar}} 
 \end{figure}

In Figs.~\ref{fig:vector} and~\ref{fig:scalar} we show the COHERENT and reactor constraints on  $\frac{g^\prime q^2}{\Lambda^2}=\frac{((g_L+g_R) g_\nu) q_0^2}{2\Lambda^2}$ at $2\sigma$ for a vector or scalar mediator, respectively,  as a function of the mediator mass. $\frac{g^\prime q^2}{\Lambda^2}$ represents the coupling strength between quarks and neutrinos as a function of energy and reduces to $g^\prime$ if there is no form factor for the coupling. We choose $q_0$ to be a typical momentum for the experiment, e.g., $q_0=50$ MeV and 30 MeV are used for COHERENT and reactor experiments, respectively. To compare with the limits for the case without a form factor, we plot the corresponding limits using dashed lines. The quarks may have direct couplings to the $\Z$ and may also couple via  DM loops in a given model, in which case  the solid and dashed lines must be combined to obtain constraints on the couplings. The plateau for small mediator masses arises because $m^{\prime2} \ll q^2$ which makes the limits independent of the mediator mass. In the regime of  large mediator masses, the slope of the limit curves is 2 since the effective couplings become $\frac{g\prime}{m^{\prime2}}$, i.e.,  $\log g^\prime \propto 2\log m^\prime$. Also notice that there is a bump in the low mass region for future COHERENT and reactor experiments because a combination of the form factor and the mediator propagator yields $\frac{q^2}{q^2+m^{\prime 2}}\sim1$, so that the mediator-induced spectral distortion is suppressed. On the other hand, for the case with no form factor, the shape distortion persists for low masses, which makes the limits stronger compared to the $F\left(q^2\right)\sim q^2$ case. Note that  direct detection constraints are  nonexistent for sub-GeV DM and collider bounds are nonexistent for a GeV mediator which allows  a lot of the parameter space to be unconstrained for $g\leq 1$.

\begin{figure}[t]
   \includegraphics[width=7.5cm]{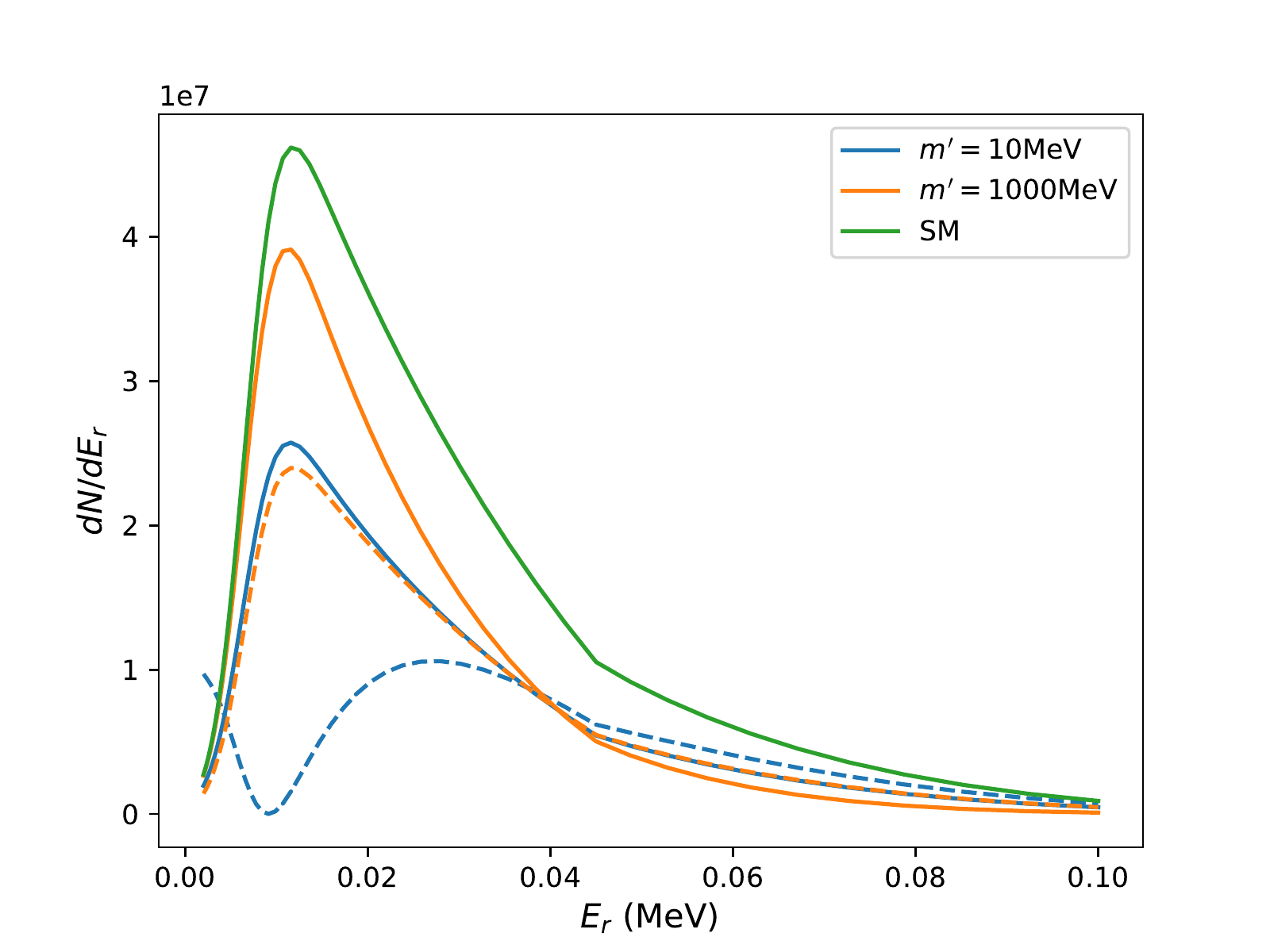}\includegraphics[width=7.5cm]{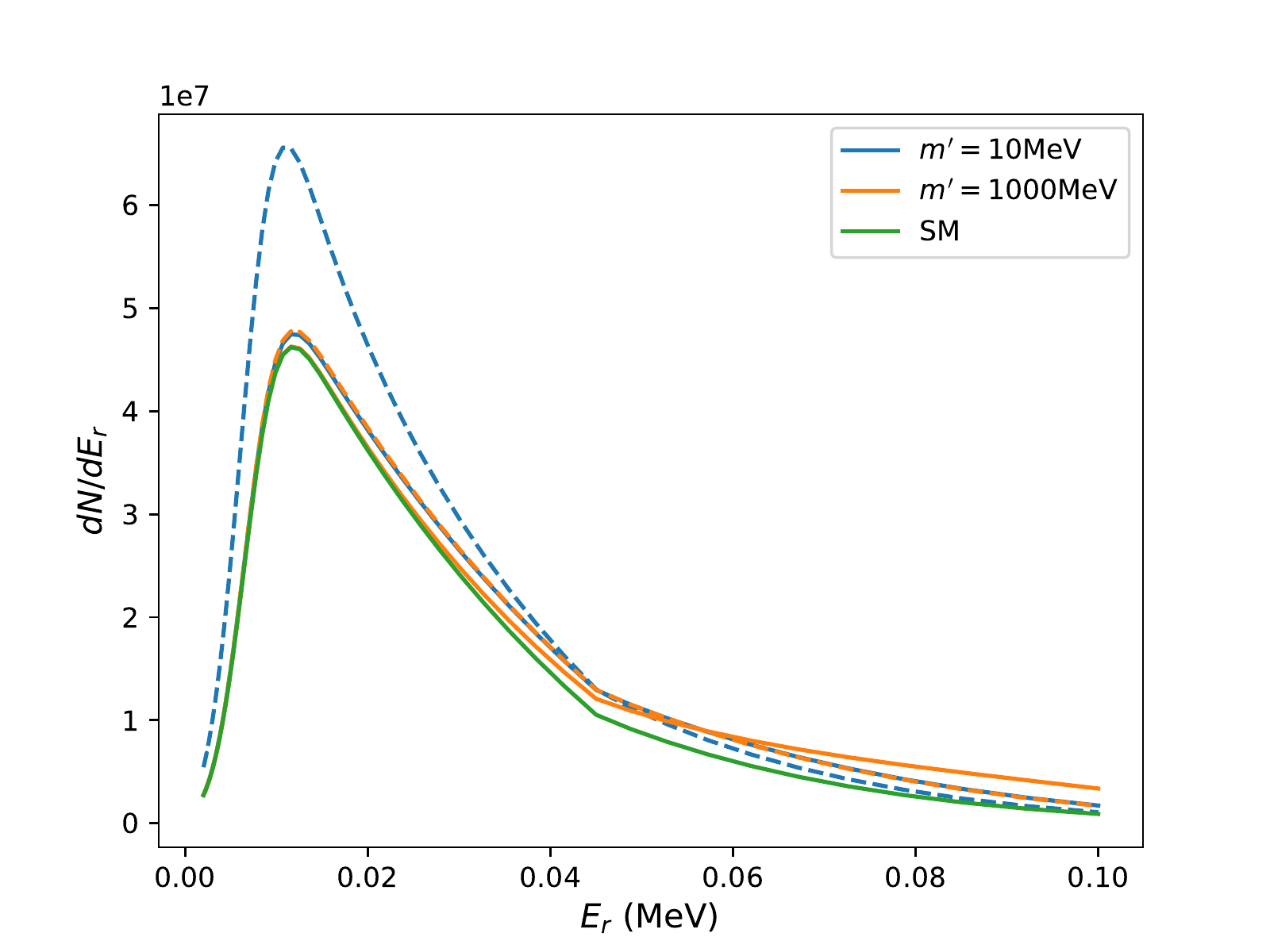}\\ 
   \caption{Spectrum of neutrino scattering off Ar detector with 1 ton$-$year exposure, with $\Lambda=100$~MeV. The left panel is for the vector mediator and the right panel is for the scalar mediator. Here the couplings for non-standard interactions are taken from the bound of current COHERENT CsI limit. The dashed lines show the spectrum without a form factor.\label{fig:sns_spect}}
 \end{figure}

An effect of the form factor, $F(q^2)\sim q^2$, is that the spectral shapes  differ from the SM prediction and from new physics models  with  $F(q^2)=1$. 
To illustrate this, we show the spectrum of coherent scattering off a Ar target in Fig.~\ref{fig:sns_spect}. 
We choose the coupling $g$ from current COHERENT constraints for $F(q^2)\sim q^2$ (solid line) and $F(q^2)=1$ (dashed line).
The main difference between the solid lines and dashed lines are at the higher energy end because the form factor $q^2$ enhances the deviation from the SM.
At low energy, the spectrum is suppressed by the detection efficiency.

\subsection{Solar neutrinos} 
Several solar neutrino experiments, for example Super-K~\cite{Abe:2010hy}, SNO~\cite{Aharmim:2011vm}, and Borexino~\cite{Agostini:2017ixy}, are sensitive to the neutrino-electron elastic scattering detection channel. 
Since the typical momentum transfer that solar neutrino experiments are sensitive to is $\sim 0.4$~MeV, it is possible to probe much smaller values of $\Lambda$ as compared to reactor and accelerator CE$\nu$NS experiments. 
Here we consider all the most prominent low energy components of the solar neutrino flux that Borexino is sensitive to, i.e., $pp$, $pep$, and $^7$Be. We choose the high metallicity solar model as defined in Ref.~\cite{Robertson:2012ib} for our baseline Standard Solar Model (SSM), and comment on the impact of the model uncertainties below. 

For solar neutrino experiments, the systematic uncertainties dominate. So we define $\chi^2$ for each component of the solar flux to be
\begin{equation}
  \chi^2 = \frac{\left(N_{th}-N_{obs}\right)^2}{N_{obs}\sigma}\,,
\end{equation}
where $\sigma$ is the percent uncertainty in the measurement (including experimental and theoretical uncertainties in quadrature) with $\sigma_{pp}=0.11$, $\sigma_{^7Be}=0.03$, and $\sigma_{pep}=0.21$~\cite{Agostini:2017ixy}. To obtain a combined limit we define $\chi^2=\chi^2_{pp}+\chi^2_{^7Be}+\chi^2_{pep}$.

\begin{figure}[t]
  \centering\includegraphics[width=12.5cm]{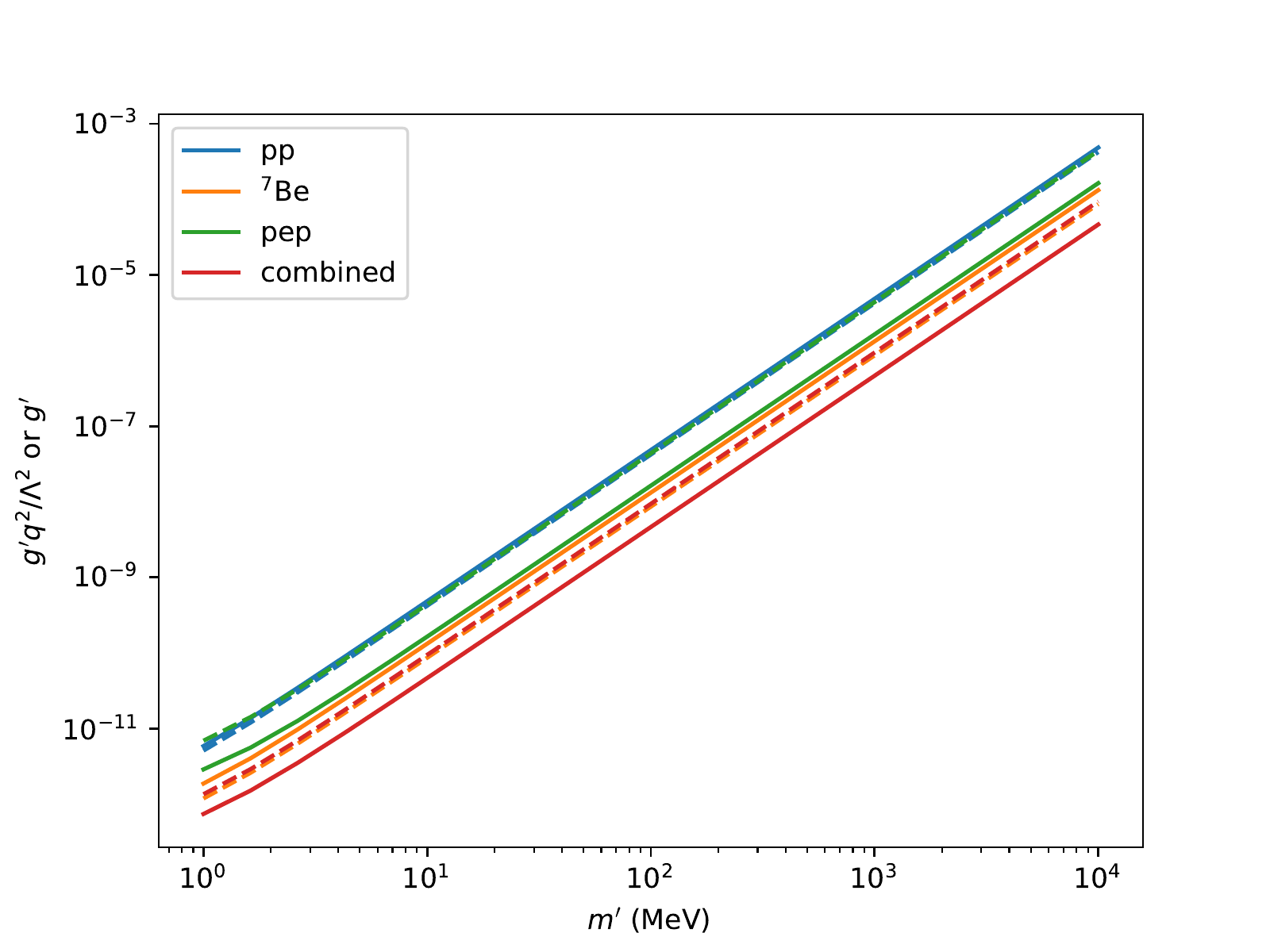}~~\\
   \caption{Constraints at $2\sigma$ from the Borexino experiment on a vector mediator with $F\left(q^2\right)\sim q^2$ as a function of the mediator mass, compared to the case of a mediator without a form factor (dashed line). We set $q=0.5$~MeV and $\Lambda=10$~MeV for the form factor case to compare it to the no-form-factor case.\label{fig:solar_vector}}
 \end{figure}

\begin{figure}[t]
\label{coherent_scalar}
  \centering\includegraphics[width=12.5cm]{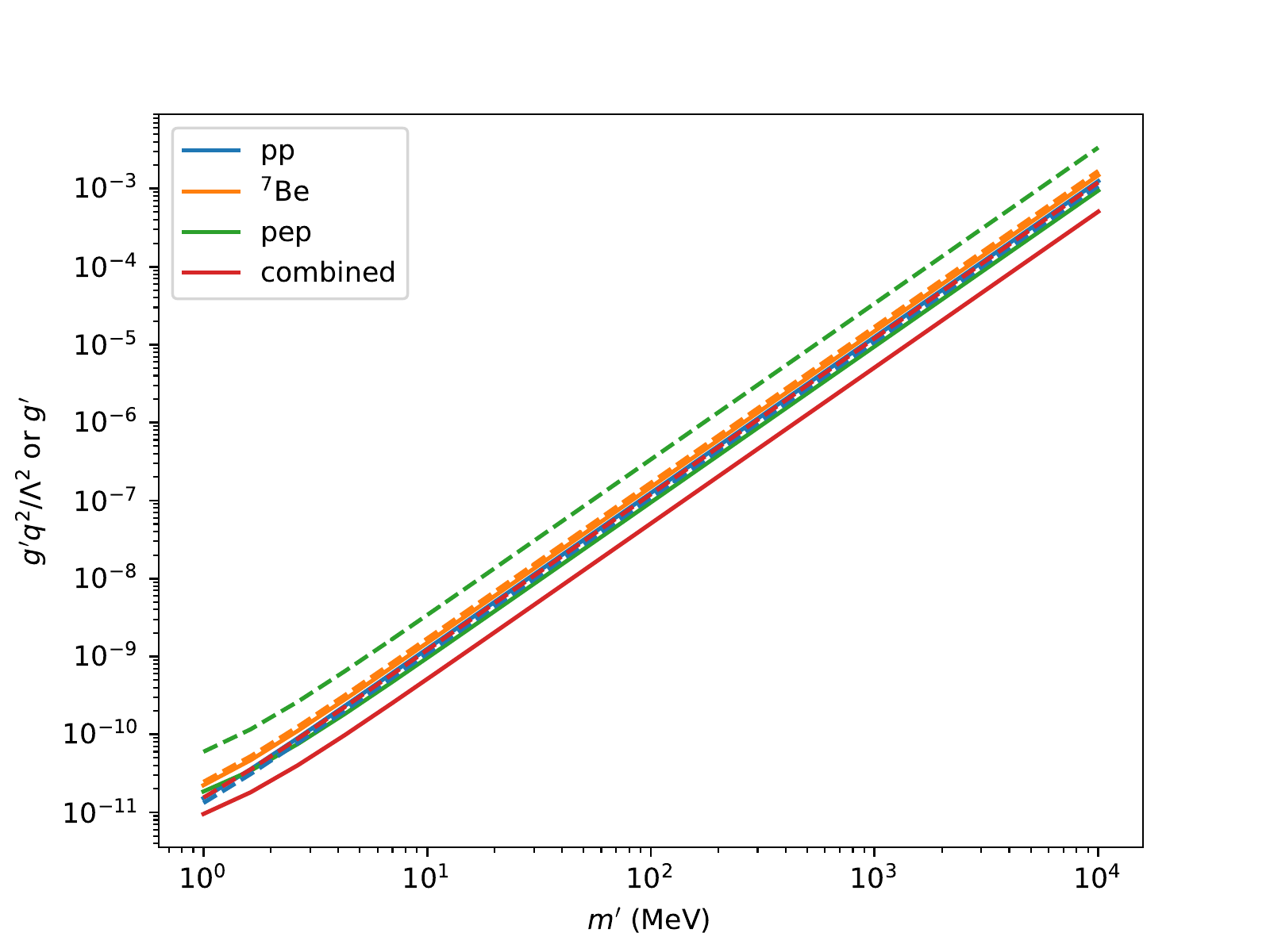}~~\\
   \caption{Constraints at $2\sigma$ from the Borexino experiment on a scalar mediator with $F\left(q^2\right)\sim q^2$ as a function of the mediator mass, compared to the case of a mediator without a form factor (dashed line). We set $q=0.5$~MeV and $\Lambda=10$~MeV for the form factor case to compare it to the no-form-factor case.\label{fig:solar_scalar}} 
 \end{figure}

 In Figs.~\ref{fig:solar_vector} and~\ref{fig:solar_scalar}, we show the constraints on the $ee\nu\nu$ coupling from Borexino~\cite{Agostini:2017ixy}. We find that the $pp$ and $^7Be$ components provide the strongest constraints on $F(q^2)\sim q^2$ because of their higher event rates and smaller flux uncertainties. This is despite the fact that the $pep$ component has larger
 spectral distortions (for the form-factor case relative to the $F(q^2)=1$ case) due to its higher energy. The limit plots are  valid as long as $\Lambda^2\gg q^2$.
 
\begin{figure}[htb!]
   \includegraphics[width=7.5cm]{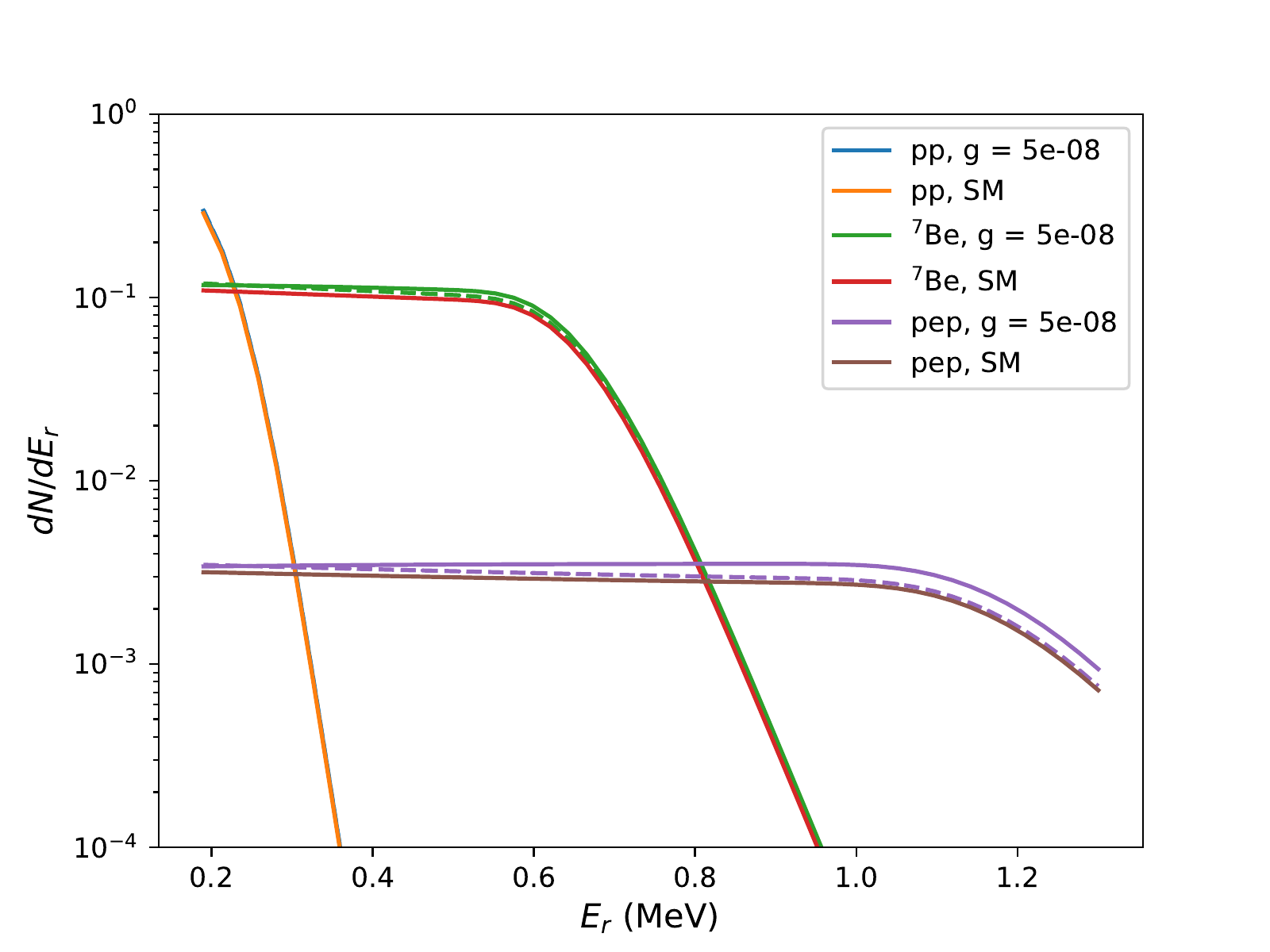}\includegraphics[width=7.5cm]{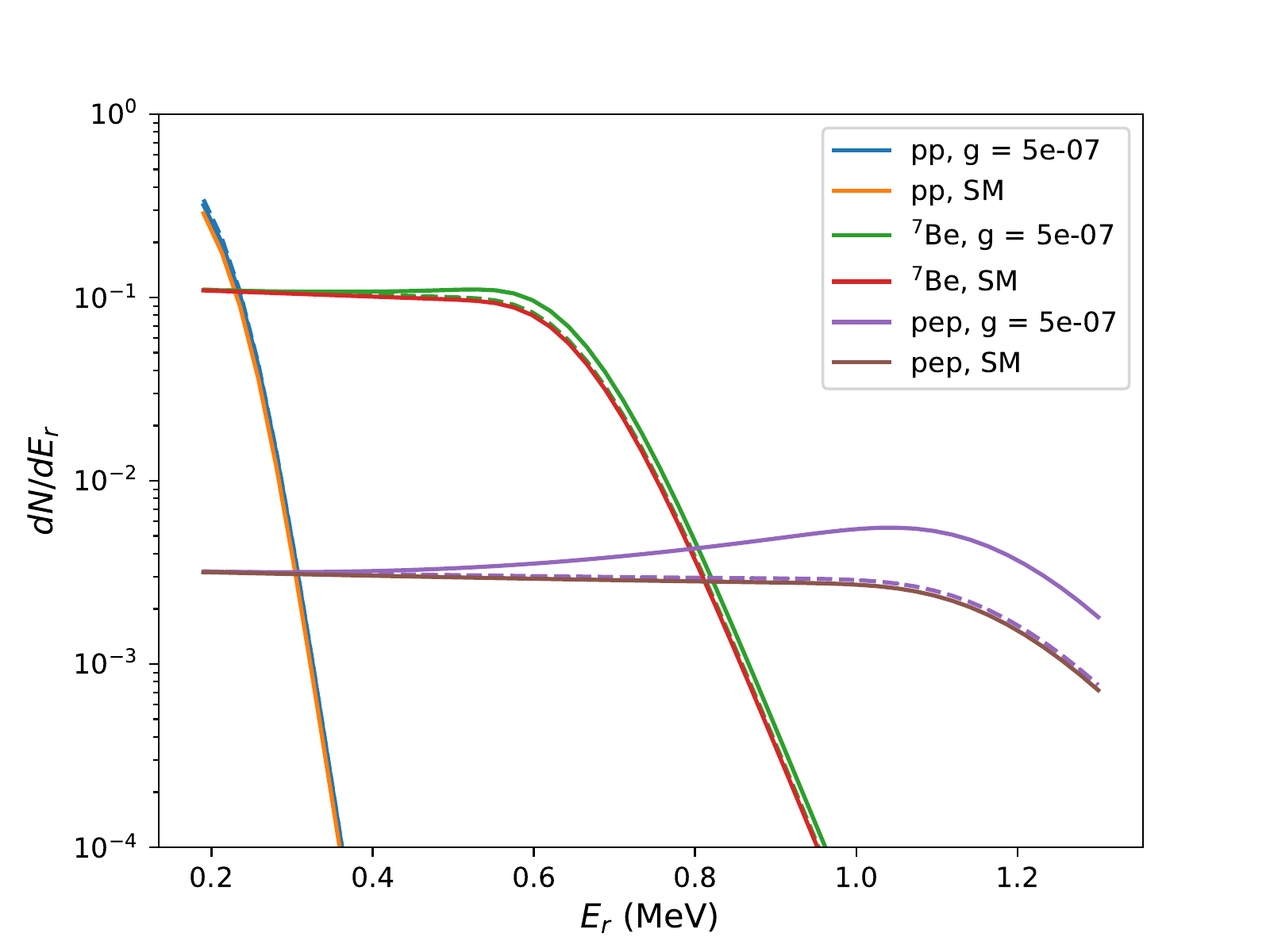}\\ 
   \caption{Spectra of solar neutrino scattering off electrons, with $m^\prime=10$~MeV and $\Lambda=10$~MeV, and scaled to match the Borexino measurement. The left panel is for the vector mediator and the right panel is for the scalar mediator. Dashed lines are the spectra without a form factor. To make a fair comparison, for the latter case we scale $g$ by a factor of $q^2/\Lambda^2$.\label{fig:solar_spect}}
 \end{figure}

 As for the nucleus scattering case, the recoil spectra in Fig. \ref{fig:solar_spect} show that the $F(q^2)\sim q^2$ case is different from the $F(q^2)=1$ case.
 We see that the major differences in the spectra are at high energies.
 The differences for the scalar case are more significant than for the vector case because in the vector scenario the $q^2$ enhancement is suppressed by the interference between SM and new physics contributions.
 
\section{B anomalies} 
 In the SM the three families of quarks and leptons are identical except for their masses. Tests of the universality of leptonic interactions are crucial probes of new physics. Recently,
hints of lepton universality violating (LUV) measurements in $B$ decays have attracted a lot of attention. These anomalies are found in the charged current $ \bctaunutau$ and neutral current $\bsll$ transitions.
Here we focus on the neutral current anomalies though the anomalies might be related~ \cite{datta_shiv, isidori1, criv1}. 
The LHCb Collaboration has measured the
ratio $R_{K^*} \equiv {\cal B}(B^0 \to K^{*0} \mu^+ \mu^-)/{\cal
  B}(B^0 \to K^{*0} e^+ e^-)$ in two ranges of the dilepton
invariant mass-squared $q^2$~\cite{Aaij:2017vbb}:
\beq
R_{K^*}^\expt = 
\left\{ 
\begin{array}{cc}
0.660^{+0.11}_{-0.07}~{\rm (stat)} \pm 0.03~{\rm (syst)} ~,~~ & 0.045 \le q^2 \le 1.1 ~{\rm GeV}^2 ~, \ \ \ (\rm{low} \ q^2)\\
0.69^{+0.11}_{-0.07}~{\rm (stat)} \pm 0.05~{\rm (syst)} ~,~~ & 1.1 \le q^2 \le 6.0 ~{\rm GeV}^2 ~, \ \ \ (\rm{central}\ q^2)\,.
\end{array}
\right.
\eeq
These differ from the SM
by 2.2-2.4$\sigma$ (low $q^2$) and 2.4-2.5$\sigma$
(central $q^2$), which supports the hint  of lepton
nonuniversality observed earlier in a similar ratio with $K$ mesons. The observable in this case is $R_K \equiv {\cal B}(B^+ \to K^+ \mu^+ \mu^-)/{\cal
  B}(B^+ \to K^+ e^+ e^-)$~\cite{hiller1, hiller2}, which  was measured by
 LHCb~\cite{RKexpt}:
\beq
R_K^\expt = 0.745^{+0.090}_{-0.074}~{\rm (stat)} \pm 0.036~{\rm (syst)} ~, \ \ \ \ 1 \le q^2 \le 6.0 ~{\rm GeV}^2\,,
\label{RKexpt}
\eeq
and found to differ  from the SM prediction, $R_K^\SM = 1 \pm 0.01$~\cite{IsidoriRK} by $2.6\sigma$.  
Other anomalies also appear in the branching ratios and angular observables of certain $ b \to s \mu^+ \mu^-$ decays.
While many new physics models with new heavy states have been discussed to address these anomalies, it was pointed out that new physics with light mediators could also explain these anomalies~\cite{ZMeV}.
In particular, with heavy new physics it is difficult to understand the $R_{K^*}$ measurement  in the low $q^2$ bin,
$0.045 \le q^2 \le 1.1~{\rm GeV}^2$, along with the measurement  of $R_{K^*}$ in the central  $q^2$ bin and the measurement of $R_K$.

For light new physics in the MeV range a resolution of the $R_K$ and $R_{K^*}$ measurements in the central $q^2$ bin along with other angular observables in $\bsmumu$ decays is possible with the light states coupling only to muons \cite{ZMeV, datta_rk, SZMEV}. 
In addition, in this framework the discrepancy in the anomalous magnetic moment of the muon can also be explained and there are interesting implications for nonstandard neutrino interactions. However, the measurement of $\RKstar$ in the low $q^2$ bin cannot be satisfactorily explained.  
For the model to work  a nontrivial form factor for the flavor changing $ bs X$ vertex is required, where $X$ is a light state. This can happen if the  $bsX$ coupling is induced  at loop level due to some additional light  new physics \cite{SZMEV} just as we have
considered in the case of neutrino scattering.
To explain  the $\RK$ and $\RKstar$ in all bins with a light mediator is difficult and requires $X$ to couple to electrons rather than muons \cite{SZMEV}. In this case the anomalies in the angular observables in $\bsmumu$ decays remain unexplained. This might suggest that there is different new physics responsible  for measurements in different $q^2$ bins.
 One can also aim to understand only the low $q^2$ bin  $\RKstar$ measurement and such an approach is followed in Ref.~\cite{alt}. 

It is possible to  connect $B$  decays to coherent neutrino scattering by generalizing  Eq.~(\ref{lagZprime}) to include all generations of quarks. We write the modified Lagrangian as
 \bea
\cal{L} & =&\frac{g}{\Lambda^2} \bar{q'}_i \gamma^{\mu} P_{L,R}Y^{ \prime i,j}_{U,D} q'_j \bar{\chi}\gamma_\mu ( 1 \pm \gamma_5) \chi +i  \bar{\chi} \gamma^\nu \left[ \partial_\nu -i g_{\chi}    Z^{\prime \nu}\right]\chi - m_{\chi} \bar{\chi}{\chi} + \frac{1}{2}m_{Z^{\prime}}^2 Z_{\mu}^{\prime}Z^{ \prime \mu} \nonumber\\
&=& H_{eff} + J_{\mu, \chi} Z^{\prime, \mu} +i  \bar{\chi} \gamma^\nu  \partial_\nu \chi - m_{\chi} \bar{\chi}{\chi} + \frac{1}{2}m_{Z^{\prime}}^2 Z_{\mu}^{\prime}Z^{ \prime \mu} , \
\label{lagZprimegeneral}
\eea
where $i,j$ are the family indices and $Y_{U,D}$ are the flavor couplings for the up and down quarks. To simplify the discussion we assume that only the left-handed  quarks are involved in the interactions with the $\chi$ fields. 
However, in order to satisfy the $\RK$ and $\RKstar$ anomalies we need flavor violation in the $b-s$ sector arising from the following Yukawa matrices:
\bea
Y_D &=&
\left(
\begin{array}{ccc}
g_{1} & 0 & 0 \\
0 & a_1 & b_1 \\
0 & b_1 &  c_1\end{array}
\right) , \nonumber\\
Y_U  & = & V_{CKM}Y_DV_{CKM}^{\dagger}\,,
\eea

The origin of the texture in the $Y_{U,D}$ can be understood by introducing a new gauge symmetry motivated by a $U(1)_{\mu-\tau}$ 
model~\cite{Lam:2001fb,Kitabayashi:2002jd,Grimus:2003kq,Koide:2003rx}, and including the quark sector. We assume that the Lagrangian has a similar symmetry in the quark sector with the following new Yuakawa terms: $\lambda^d_1 \bar{Q}_L^{(2)}\tilde{H}_3D_R^{(3)}+\lambda_2 \bar{Q}_L^{(3)}\tilde{H}_4D_R^{(2)}$, where $\tilde{H}_{3,4}$ have new gauge charges 2a,  -2a respectively, in addition to the  SM weak charge assignments $(2, 1/2)$ under $SU(2)_L$ and $U(1)_Y$. Similar terms for the up sector are present as well. Such a model has been constructed in Ref.~\cite{Datta:2005fa}. Here we assume that the quarks transform as $(0, a, -a)$ but we could have assumed $(a, a, -2a)$ as well with different charge assignments for the new Higgs.

 In the weak interaction  basis,
the couplings to $Z^\prime$ associated with the new symmetry is diagonal,
\beq
Y_{U,D}^{\prime} =
\left(
\begin{array}{ccc}
g_1 & 0 & 0 \\
0 & g_2 & 0 \\
0 & 0 & g_3
\end{array}
\right)\,.
\eeq
In transforming
from the gauge basis to the mass basis (with the contributions arising from the off-diagonal terms in the Lagrangian), we write
\beq
u'_L = U_L u_L ~,~~ d'_L = D_L d_L ~,~~ 
\label{transformations}
\eeq
where $U_{L}$ and $D_{L}$  are $3\times 3$ unitary matrices  for the up and down quarks respectively, and the
spinors $u^{(\prime)}$ and  $d^{(\prime)}$ include all three generations of fermions. 
The CKM matrix is given by $V_{CKM}=U_L^\dagger D_L$ and
after transforming to the mass basis we can rewrite Eq.~(\ref{lagZprimegeneral})  with the $Y_{U,D}^{\prime}$ matrices replaced by  
$Y_{D}= D_L^{\dagger} Y_D^{\prime}D_L$ and
$Y_{U}= U_L^{\dagger} Y_U^{\prime}U_L$.
Now if all the mixing is in the up sector with $D_L=I$ then there is no FCNC in the down sector. To generate the $ b \to s$ transition we
use  the
assumption of Ref.~\cite{datta2}  that $D_{L}$
involves only the second and third generations:
\beq
D_L =
\left(
\begin{array}{ccc}
1 & 0 & 0 \\
0 & \cos\theta_D & \sin\theta_D \\
0 & -\sin\theta_D & \cos\theta_D
\end{array}
\right)\,.
\eeq
The $Y_{U,D}$ matrices then have the explicit form,
\bea
Y_D &=&
\left(
\begin{array}{ccc}
g_{1} & 0 & 0 \\
0 & c_D^2 g_{2} + s_D^2 g_{3} & c_Ds_D( g_{2}-g_{3}) \\
0 & c_Ds_D( g_{2}-g_{3}) &  c_D^2 g_{3} + s_D^2 g_{2}\end{array}
\right) , \nonumber\\
Y_U  & = & V_{CKM}Y_DV_{CKM}^{\dagger}\,,
\eea
where $c_D \equiv \cos \theta_{D}$ and    $s_D \equiv \sin \theta_{D}$. We see that  in the down sector flavor changing $b \to s$ transitions occur with coupling
$g_{bs}= c_Ds_D( g_{2}-g_{3})$. The form factor for coherent scattering is $F(q^2) = g_L \frac{q^2}{\Lambda^2}$  while for the $B$ decays it is
 $F(q^2) = g_{Lbs} \frac{q^2}{\Lambda^2}$  with $ g_L \propto g_1$, $g_{Lbs} \propto g_{bs}$ and
  $ \frac{g_1}{g_{bs}} = \frac{g_L}{g_{Lbs}}$. If all the $g_i$ are of the same order of magnitude then $g_{Lbs} < g_L$.

The breakdown of lepton flavor universality required for the $R_{K^{(\ast)}}$ anomaly can arise in $U(1)_{\mu-\tau}$ symmetry models. We now compare the flavor violating terms with the flavor conserving terms in the quark sector. 

 Combining the $B$ decay anomalies with the results from coherent scattering allows us to check for the consistency of this framework.
 We focus on the  $\Z$ models. Figure~\ref{fig:vector} 
 gives the bound on the diagonal term using the COHERENT experiment,
 \bea
\Lambda^2 > q_0^2 \frac{  g_L g_\mu   }{X_l}\,,
\label{vconstraint}
\eea
where $g_\mu$ is the $Z^\prime$ coupling to muon neutrinos, and we can read off $X_l$ from the figure. 

\begin{figure}
 	\centering\includegraphics[width=12.5cm]{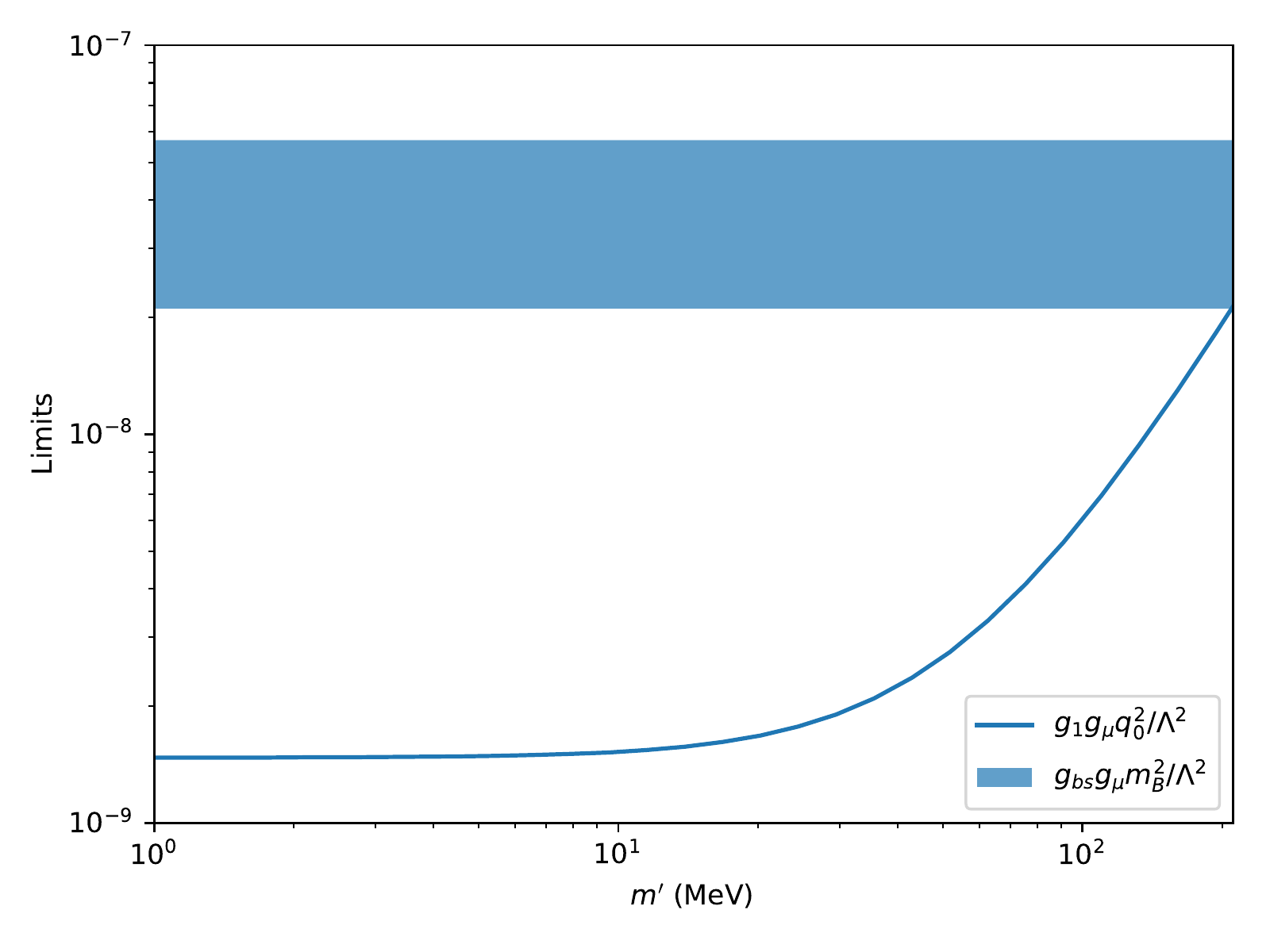}
 	\caption{Scale ($\Lambda$) independent comparison between current $2\sigma$ COHERENT and $1\sigma$ $B$ decay constraints on diagonal or off-diagonal couplings for a  mediator lighter than $2m_\mu$. \label{fig:compare}}
 \end{figure}

We now turn our attention to the $R_{K^{(\ast)}}$ anomaly which involves a flavor violating $b-s$ interaction with charged muons. Using the recent results on $R_{K^{(\ast)}}$, we obtain  a constraint on the flavor violating term. 
We assume the left handed leptons have identical couplings and so $g_{\mu}$ can be fixed from the muon anomalous magnetic moment measurement and neutrino trident production. Using $g_{\mu} \sim 10^{-3}$~\cite{datta_rk,SZMEV} and $X_l \sim 10^{-9}$ from Fig.~\ref{fig:vector} 
we obtain
\bea
\Lambda > 10^3 q_0 g_L\,,\
\eea 
which gives $ \Lambda > 50g_L $~GeV for $q_0=50$~MeV.
In $B$ decays the  relevant $q_0$ is taken to be $m_B$ and so with the additional assumption of
  $SU(2)_L$ symmetry for the left handed leptonic couplings  we obtain~\cite{SZMEV},
\bea
g_{Lbs}\frac{m_B^2}{\Lambda^2} g_\mu \sim X_h\,.
\eea  
Combining this with Eq.~(\ref{vconstraint}), we get
\bea
 \frac{g_{Lbs}}{g_{L}} = \frac{g_{bs}}{g_1} > \frac{X_h}{X_l} \frac{q_0^2}{m_B^2}\,.
 \label{vconstraint1}
 \eea
 Using $X_h \sim 10^{-8}$~\cite{SZMEV} and $X_l \sim 10^{-9}$ from Fig.~\ref{fig:compare}, we find $g_{bs}/g_1 \sim 10^{-3}$ for $m'$ between $1-10$~MeV, and $g_{bs}/g_1 \sim 10^{-4}$ for $m'=100$~MeV.
 However, if the bounds from coherent scattering get stronger, then 
  $\frac{g_{Lbs}}{g_{L}}$ will increase and lead to  tension in the framework. A similar analysis can also be done with scalar mediators where tighter constraints are obtained.
The $B$ anomalies also indicate lepton universality violating new physics which will be interesting to check in neutrino scattering. For instance if the $R_{K^{(*)}}$ anomalies are explained with
mediators coupling differently to muons and electrons then $\nu_\mu$ and $\nu_e$ scattering may show different new physics effects.


\section{Conclusions}

We have explored the limits of the effective couplings arising from a high energy scale ($\Lambda$) hidden sector associated with dark matter.
We considered two general models which give rise to a coupling form factor that is proportional to the momentum $q^2$. 
The $Z^\prime$ model corresponds to vector couplings between neutrinos and quarks, and the $S$ model corresponds to scalar couplings.
At low energies, we have shown that it is possible to probe $\Lambda$ via CE$\nu$NS experiments via the form factor which is induced by a DM ($\chi$) loop.
We considered scenarios in which  $\Lambda$ is $\geq 50$~MeV, and in which the energy scale is $\leq 1$~MeV. 
CE$\nu$NS experiments can probe the former case since $\Lambda$ is higher than the momentum transfer but these experiments are unable to probe the latter case.
To probe the scenario with small $\Lambda$, we used solar electron scattering experiments for which the momentum transfer is low.

In the $Z^\prime$ model, COHERENT constrains the  coupling to be $\sim 10^{-8}$ at $2\sigma$ for small mediator masses. For large mediator masses, the bound scales according to $\log g^\prime \propto 2\log m^\prime$, as shown in Fig.~\ref{coherent_vector}.
Atomic parity violation does better than most of the CE$\nu$NS experiments except those using reactor neutrinos.
For small $\Lambda$ the Borexino experiment puts $2\sigma$ constraints  on the couplings ${\cal O}(10^{-7})$ for a 100~MeV mass mediator. Since the momentum transfer is much smaller, the constraints scale like $\log g^\prime \propto 2\log m^\prime$ as shown in Fig.~\ref{fig:solar_vector}.

In the $S$ model, COHERENT constrains the coupling to be $\sim 10^{-9}$ for small mediator masses. 
For large mediator masses, the bound scales according to $\log g^\prime \propto 2\log m^\prime$, as shown in Fig.~\ref{coherent_scalar}.
Atomic parity violation experiments do not constrain models with scalar mediators.
For small $\Lambda$, the Borexino experiment  puts $2\sigma$ constraints on the couplings ${\cal O}(10^{-7})$ for a 100~MeV mass mediator. Since the momentum transfer is much smaller, the constraints scale like $\log g^\prime \propto 2\log m^\prime$ as shown in 
Fig.~\ref{fig:solar_scalar}. 

 Finally, we have extended our framework to quarks of all generations and have addressed the $R_K$ and $R_{K^*}$ anomalies in rare $B$ decays. We have shown that a resolution of the anomalies  consistent with the present coherent scattering data is possible but future constraints  from coherent scattering  will  provide stringent tests of the $B$ anomalies explanation.

\section*{Acknowledgements}

A.D. acknowledges the hospitality of University of California, Irvine where this work was completed. D.M. thanks the Mitchell Institute at Texas A\&M University for its support and hospitality while this work was in progress. A.D. acknowledges support from NSF under Grant No. PHY-1414345.  B.D. and L.S. acknowledge support from DOE Grant de-sc0010813. S.L. acknowledges support from TAMU - College of Science - STRP. D.M. acknowledges support from DOE Grant No.~de-sc0010504.

\bibliographystyle{JHEP}
\bibliography{main}

\providecommand{\href}[2]{#2}\begingroup\raggedright\begin{thebibliography}{10}

\bibitem{Akimov:2017ade}
{\scshape COHERENT} collaboration, D.~Akimov et~al., \emph{{Observation of
  Coherent Elastic Neutrino-Nucleus Scattering}},
  \href{https://doi.org/10.1126/science.aao0990}{\emph{Science} {\bfseries 357}
  (2017) 1123} [\href{https://arxiv.org/abs/1708.01294}{{\ttfamily
  1708.01294}}].

\bibitem{Agostini:2017ixy}
{\scshape Borexino} collaboration, M.~Agostini et~al., \emph{{First
  Simultaneous Precision Spectroscopy of $pp$, $^7$Be, and $pep$ Solar
  Neutrinos with Borexino Phase-II}},
  \href{https://arxiv.org/abs/1707.09279}{{\ttfamily 1707.09279}}.

\bibitem{SZMEV}
A.~Datta, J.~Kumar, J.~Liao and D.~Marfatia, \emph{{New light mediators for the
  $R_K$ and $R_{K^*}$ puzzles}},
  \href{https://doi.org/10.1103/PhysRevD.97.115038}{\emph{Phys. Rev.}
  {\bfseries D97} (2018) 115038}
  [\href{https://arxiv.org/abs/1705.08423}{{\ttfamily 1705.08423}}].

\bibitem{Goodman:2010ku}
J.~Goodman, M.~Ibe, A.~Rajaraman, W.~Shepherd, T.~M.~P. Tait and H.-B. Yu,
  \emph{{Constraints on Dark Matter from Colliders}},
  \href{https://doi.org/10.1103/PhysRevD.82.116010}{\emph{Phys. Rev.}
  {\bfseries D82} (2010) 116010}
  [\href{https://arxiv.org/abs/1008.1783}{{\ttfamily 1008.1783}}].

\bibitem{Bai:2010hh}
Y.~Bai, P.~J. Fox and R.~Harnik, \emph{{The Tevatron at the Frontier of Dark
  Matter Direct Detection}},
  \href{https://doi.org/10.1007/JHEP12(2010)048}{\emph{JHEP} {\bfseries 12}
  (2010) 048} [\href{https://arxiv.org/abs/1005.3797}{{\ttfamily 1005.3797}}].

\bibitem{Fan:2010gt}
J.~Fan, M.~Reece and L.-T. Wang, \emph{{Non-relativistic effective theory of
  dark matter direct detection}},
  \href{https://doi.org/10.1088/1475-7516/2010/11/042}{\emph{JCAP} {\bfseries
  1011} (2010) 042} [\href{https://arxiv.org/abs/1008.1591}{{\ttfamily
  1008.1591}}].

\bibitem{Elor:2018xku}
G.~Elor, H.~Liu, T.~R. Slatyer and Y.~Soreq, \emph{{Complementarity for Dark
  Sector Bound States}},  \href{https://arxiv.org/abs/1801.07723}{{\ttfamily
  1801.07723}}.

\bibitem{Datta:2013kja}
A.~Datta, M.~Duraisamy and D.~Ghosh, \emph{{Explaining the $B \to K^\ast \mu^+
  \mu^-$ data with scalar interactions}},
  \href{https://doi.org/10.1103/PhysRevD.89.071501}{\emph{Phys. Rev.}
  {\bfseries D89} (2014) 071501}
  [\href{https://arxiv.org/abs/1310.1937}{{\ttfamily 1310.1937}}].

\bibitem{Cerdeno:2016sfi}
D.~G. Cerdeno, M.~Fairbairn, T.~Jubb, P.~A.~N. Machado, A.~C. Vincent and
  C.~Bahm, \emph{{Physics from solar neutrinos in dark matter direct detection
  experiments}}, \href{https://doi.org/10.1007/JHEP09(2016)048,
  10.1007/JHEP05(2016)118}{\emph{JHEP} {\bfseries 05} (2016) 118}
  [\href{https://arxiv.org/abs/1604.01025}{{\ttfamily 1604.01025}}].

\bibitem{DelNobile:2013sia}
M.~Cirelli, E.~Del~Nobile and P.~Panci, \emph{{Tools for model-independent
  bounds in direct dark matter searches}},
  \href{https://doi.org/10.1088/1475-7516/2013/10/019}{\emph{JCAP} {\bfseries
  1310} (2013) 019} [\href{https://arxiv.org/abs/1307.5955}{{\ttfamily
  1307.5955}}].

\bibitem{Tanabashi:2018oca}
{\scshape Particle Data Group} collaboration, M.~Tanabashi et~al.,
  \emph{{Review of Particle Physics}},
  \href{https://doi.org/10.1103/PhysRevD.98.030001}{\emph{Phys. Rev.}
  {\bfseries D98} (2018) 030001}.

\bibitem{Nambu:1961tp}
Y.~Nambu and G.~Jona-Lasinio, \emph{{Dynamical Model of Elementary Particles
  Based on an Analogy with Superconductivity. 1.}},
  \href{https://doi.org/10.1103/PhysRev.122.345}{\emph{Phys. Rev.} {\bfseries
  122} (1961) 345}.

\bibitem{us}
D.~Aristizabal~Sierra, B.~Dutta and L.~Strigari, \emph{{In preparation}},
  {\emph{(2019)} }.

\bibitem{aguilar2016connie}
{\scshape CONNIE} collaboration, A.~Aguilar-Arevalo et~al., \emph{{The CONNIE
  experiment}}, \href{https://doi.org/10.1088/1742-6596/761/1/012057}{\emph{J.
  Phys. Conf. Ser.} {\bfseries 761} (2016) 012057}
  [\href{https://arxiv.org/abs/1608.01565}{{\ttfamily 1608.01565}}].

\bibitem{li2016neutrino}
{\scshape TEXONO} collaboration, H.~Bin~Li, \emph{{Neutrino and dark matter
  physics with sub-KeV Germanium detectors}},
  \href{https://doi.org/10.1088/1742-6596/718/6/062036}{\emph{J. Phys. Conf.
  Ser.} {\bfseries 718} (2016) 062036}.

\bibitem{agnolet2017background}
{\scshape MINER} collaboration, G.~Agnolet et~al., \emph{{Background Studies
  for the MINER Coherent Neutrino Scattering Reactor Experiment}},
  \href{https://doi.org/10.1016/j.nima.2017.02.024}{\emph{Nucl. Instrum. Meth.}
  {\bfseries A853} (2017) 53}
  [\href{https://arxiv.org/abs/1609.02066}{{\ttfamily 1609.02066}}].

\bibitem{Akimov:2018ghi}
{\scshape COHERENT} collaboration, D.~Akimov et~al., \emph{{COHERENT 2018 at
  the Spallation Neutron Source}},
  \href{https://arxiv.org/abs/1803.09183}{{\ttfamily 1803.09183}}.

\bibitem{an2017improved}
{\scshape Daya Bay} collaboration, F.~P. An et~al., \emph{{Improved Measurement
  of the Reactor Antineutrino Flux and Spectrum at Daya Bay}},
  \href{https://doi.org/10.1088/1674-1137/41/1/013002}{\emph{Chin. Phys.}
  {\bfseries C41} (2017) 013002}
  [\href{https://arxiv.org/abs/1607.05378}{{\ttfamily 1607.05378}}].

\bibitem{Abe:2010hy}
{\scshape Super-Kamiokande} collaboration, K.~Abe et~al., \emph{{Solar neutrino
  results in Super-Kamiokande-III}},
  \href{https://doi.org/10.1103/PhysRevD.83.052010}{\emph{Phys. Rev.}
  {\bfseries D83} (2011) 052010}
  [\href{https://arxiv.org/abs/1010.0118}{{\ttfamily 1010.0118}}].

\bibitem{Aharmim:2011vm}
{\scshape SNO} collaboration, B.~Aharmim et~al., \emph{{Combined Analysis of
  all Three Phases of Solar Neutrino Data from the Sudbury Neutrino
  Observatory}}, \href{https://doi.org/10.1103/PhysRevC.88.025501}{\emph{Phys.
  Rev.} {\bfseries C88} (2013) 025501}
  [\href{https://arxiv.org/abs/1109.0763}{{\ttfamily 1109.0763}}].

\bibitem{Robertson:2012ib}
W.~C. Haxton, R.~G. Hamish~Robertson and A.~M. Serenelli, \emph{{Solar
  Neutrinos: Status and Prospects}},
  \href{https://doi.org/10.1146/annurev-astro-081811-125539}{\emph{Ann. Rev.
  Astron. Astrophys.} {\bfseries 51} (2013) 21}
  [\href{https://arxiv.org/abs/1208.5723}{{\ttfamily 1208.5723}}].

\bibitem{datta_shiv}
B.~Bhattacharya, A.~Datta, D.~London and S.~Shivashankara, \emph{{Simultaneous
  Explanation of the $R_K$ and $R(D^{(*)})$ Puzzles}},
  \href{https://doi.org/10.1016/j.physletb.2015.02.011}{\emph{Phys. Lett.}
  {\bfseries B742} (2015) 370}
  [\href{https://arxiv.org/abs/1412.7164}{{\ttfamily 1412.7164}}].

\bibitem{isidori1}
A.~Greljo, G.~Isidori and D.~Marzocca, \emph{{On the breaking of Lepton Flavor
  Universality in B decays}},
  \href{https://doi.org/10.1007/JHEP07(2015)142}{\emph{JHEP} {\bfseries 07}
  (2015) 142} [\href{https://arxiv.org/abs/1506.01705}{{\ttfamily
  1506.01705}}].

\bibitem{criv1}
A.~Crivellin, D.~Muller and T.~Ota, \emph{{Simultaneous explanation of
  R(D$^{(*)}$) and $b\rightarrow s\mu^{+}$ $\mu^{-}$: the last scalar
  leptoquarks standing}},
  \href{https://doi.org/10.1007/JHEP09(2017)040}{\emph{JHEP} {\bfseries 09}
  (2017) 040} [\href{https://arxiv.org/abs/1703.09226}{{\ttfamily
  1703.09226}}].

\bibitem{Aaij:2017vbb}
{\scshape LHCb} collaboration, R.~Aaij et~al., \emph{{Test of lepton
  universality with $B^{0} \rightarrow K^{*0}\ell^{+}\ell^{-}$ decays}},
  \href{https://doi.org/10.1007/JHEP08(2017)055}{\emph{JHEP} {\bfseries 08}
  (2017) 055} [\href{https://arxiv.org/abs/1705.05802}{{\ttfamily
  1705.05802}}].

\bibitem{hiller1}
G.~Hiller and F.~Kruger, \emph{{More model-independent analysis of $b \to s$
  processes}}, \href{https://doi.org/10.1103/PhysRevD.69.074020}{\emph{Phys.
  Rev.} {\bfseries D69} (2004) 074020}
  [\href{https://arxiv.org/abs/hep-ph/0310219}{{\ttfamily hep-ph/0310219}}].

\bibitem{hiller2}
G.~Hiller and M.~Schmaltz, \emph{{$R_K$ and future $b \to s \ell \ell$ physics
  beyond the standard model opportunities}},
  \href{https://doi.org/10.1103/PhysRevD.90.054014}{\emph{Phys. Rev.}
  {\bfseries D90} (2014) 054014}
  [\href{https://arxiv.org/abs/1408.1627}{{\ttfamily 1408.1627}}].

\bibitem{RKexpt}
{\scshape LHCb} collaboration, R.~Aaij et~al., \emph{{Test of lepton
  universality using $B^{+}\rightarrow K^{+}\ell^{+}\ell^{-}$ decays}},
  \href{https://doi.org/10.1103/PhysRevLett.113.151601}{\emph{Phys. Rev. Lett.}
  {\bfseries 113} (2014) 151601}
  [\href{https://arxiv.org/abs/1406.6482}{{\ttfamily 1406.6482}}].

\bibitem{IsidoriRK}
M.~Bordone, G.~Isidori and A.~Pattori, \emph{{On the Standard Model predictions
  for $R_K$ and $R_{K^*}$}},
  \href{https://doi.org/10.1140/epjc/s10052-016-4274-7}{\emph{Eur. Phys. J.}
  {\bfseries C76} (2016) 440}
  [\href{https://arxiv.org/abs/1605.07633}{{\ttfamily 1605.07633}}].

\bibitem{ZMeV}
A.~Datta, J.~Liao and D.~Marfatia, \emph{{A light $Z^\prime$ for the $R_K$
  puzzle and nonstandard neutrino interactions}},
  \href{https://doi.org/10.1016/j.physletb.2017.02.058}{\emph{Phys. Lett.}
  {\bfseries B768} (2017) 265}
  [\href{https://arxiv.org/abs/1702.01099}{{\ttfamily 1702.01099}}].

\bibitem{datta_rk}
A.~K. Alok, B.~Bhattacharya, A.~Datta, D.~Kumar, J.~Kumar and D.~London,
  \emph{{New Physics in $b \to s \mu^+ \mu^-$ after the Measurement of
  $R_{K^*}$}}, \href{https://doi.org/10.1103/PhysRevD.96.095009}{\emph{Phys.
  Rev.} {\bfseries D96} (2017) 095009}
  [\href{https://arxiv.org/abs/1704.07397}{{\ttfamily 1704.07397}}].

\bibitem{alt}
W.~Altmannshofer, M.~J. Baker, S.~Gori, R.~Harnik, M.~Pospelov, E.~Stamou
  et~al., \emph{{Light resonances and the low-q$^{2}$ bin of $ {R}_{K^{*}} $}},
  \href{https://doi.org/10.1007/JHEP03(2018)188}{\emph{JHEP} {\bfseries 03}
  (2018) 188} [\href{https://arxiv.org/abs/1711.07494}{{\ttfamily
  1711.07494}}].

\bibitem{Lam:2001fb}
C.~S. Lam, \emph{{A 2-3 symmetry in neutrino oscillations}},
  \href{https://doi.org/10.1016/S0370-2693(01)00465-8}{\emph{Phys. Lett.}
  {\bfseries B507} (2001) 214}
  [\href{https://arxiv.org/abs/hep-ph/0104116}{{\ttfamily hep-ph/0104116}}].

\bibitem{Kitabayashi:2002jd}
T.~Kitabayashi and M.~Yasue, \emph{{S(2L) permutation symmetry for left-handed
  mu and tau families and neutrino oscillations in an SU(3)-L x SU(1)-N gauge
  model}}, \href{https://doi.org/10.1103/PhysRevD.67.015006}{\emph{Phys. Rev.}
  {\bfseries D67} (2003) 015006}
  [\href{https://arxiv.org/abs/hep-ph/0209294}{{\ttfamily hep-ph/0209294}}].

\bibitem{Grimus:2003kq}
W.~Grimus and L.~Lavoura, \emph{{A Discrete symmetry group for maximal
  atmospheric neutrino mixing}},
  \href{https://doi.org/10.1016/j.physletb.2003.08.032}{\emph{Phys. Lett.}
  {\bfseries B572} (2003) 189}
  [\href{https://arxiv.org/abs/hep-ph/0305046}{{\ttfamily hep-ph/0305046}}].

\bibitem{Koide:2003rx}
Y.~Koide, \emph{{Universal texture of quark and lepton mass matrices with an
  extended flavor 2 <---> 3 symmetry}},
  \href{https://doi.org/10.1103/PhysRevD.69.093001}{\emph{Phys. Rev.}
  {\bfseries D69} (2004) 093001}
  [\href{https://arxiv.org/abs/hep-ph/0312207}{{\ttfamily hep-ph/0312207}}].

\bibitem{Datta:2005fa}
A.~Datta and P.~J. O'Donnell, \emph{{The 2-3 symmetry: Flavor changing b, tau
  decays and neutrino mixing}},
  \href{https://doi.org/10.1103/PhysRevD.72.113002}{\emph{Phys. Rev.}
  {\bfseries D72} (2005) 113002}
  [\href{https://arxiv.org/abs/hep-ph/0508314}{{\ttfamily hep-ph/0508314}}].

\bibitem{datta2}
B.~Bhattacharya, A.~Datta, J.-P. Gu{\'e}vin, D.~London and R.~Watanabe,
  \emph{{Simultaneous Explanation of the $R_K$ and $R_{D^{(*)}}$ Puzzles: a
  Model Analysis}}, \href{https://doi.org/10.1007/JHEP01(2017)015}{\emph{JHEP}
  {\bfseries 01} (2017) 015}
  [\href{https://arxiv.org/abs/1609.09078}{{\ttfamily 1609.09078}}].

\end{thebibliography}\endgroup
\end{document}